\documentclass[longauth]{aa} 
\usepackage[utf8]{inputenc}

\usepackage{graphicx}
\usepackage{siunitx}
\usepackage{amsmath}
\usepackage{textcomp}
\usepackage{multirow}                
\usepackage{booktabs}                              
\usepackage{multicol,lipsum}
\usepackage{colortbl}

\usepackage{txfonts}
\usepackage[]{hyperref}
\usepackage{hyperref}
 \hypersetup{
     colorlinks=true,
     linkcolor=blue,
     filecolor=blue,
     citecolor =blue,      
     urlcolor=blue,
     }

\DeclareRobustCommand{\ion}[2]{%
\relax\ifmmode
\ifx\testbx\f@series
{\mathbf{#1\,\mathsc{#2}}}\else
{\mathrm{#1\,\mathsc{#2}}}\fi
\else\textup{#1\,{\mdseries\textsc{#2}}}%
\fi}

\usepackage{comment}
\usepackage{balance}
\usepackage{subfig}

\begin{document}

   \title{The GAPS programme at TNG LII. Spot modeling of V1298 Tau using SpotCCF tool. \thanks{Based on observations made with the Italian Telescopio Nazionale
Galileo (TNG) operated by the Fundación Galileo Galilei (FGG) of the
Istituto Nazionale di Astrofisica (INAF) at the Observatorio del Roque
de los Muchachos (La Palma, Canary Islands, Spain).}}

   \author{C. Di Maio
          \inst{1}
          \and
          A. Petralia\inst{1}
          \and
          G. Micela\inst{1}
          \and
          A. F. Lanza\inst{2}
          \and
          M. Rainer\inst{3} 
          \and 
          L. Malavolta\inst{4} 
          \and 
          S. Benatti\inst{1}
          \and
          L. Affer\inst{1}
          \and
          J. Maldonado\inst{1}
          \and
          S. Colombo\inst{1}
          \and
          M. Damasso\inst{5} 
          \and
          A. Maggio\inst{1}
          \and
          K. Biazzo\inst{6} 
          \and 
          A. Bignamini\inst{7} 
          \and 
          F. Borsa\inst{3} 
          \and 
          W. Boschin\inst{8,9,10} 
          \and
          L. Cabona\inst{11} 
          \and 
          M. Cecconi\inst{8} 
          \and 
          R. Claudi\inst{11} 
          \and 
          E. Covino\inst{12} 
          \and 
          L. Di Fabrizio\inst{8} 
          \and 
          R. Gratton\inst{11} 
          \and 
          V. Lorenzi\inst{8,9} 
          \and 
          L. Mancini\inst{13,5,14} 
          \and 
          S. Messina\inst{2} 
          \and 
          E. Molinari\inst{15} 
          \and 
          M. Molinaro\inst{7} 
          \and 
          D. Nardiello\inst{4,11} 
          \and 
          E. Poretti\inst{3,8} 
          \and 
          A. Sozzetti\inst{5} 
          } 

   \institute{INAF – Osservatorio Astronomico di Palermo, Piazza del Parlamento, 1, 90134 Palermo, Italy.\\
              \email{claudia.dimaio@inaf.it}
        \and
            INAF – Osservatorio Astrofisico di Catania, Via S. Sofia 78, 95123 Catania, Italy
        \and
            INAF – Osservatorio Astronomico di Brera, Via E. Bianchi 46, 23807 Merate (LC), Italy
        \and 
             Dipartimento di Fisica e Astronomia "Galileo Galilei", Università di Padova, Vicolo dell'Osservatorio 3, I-35122 Padova, Italy
        \and 
            INAF – Osservatorio Astrofisico di Torino, Via Osservatorio 20, I-10025, Pino Torinese (TO), Italy
        \and
            INAF – Osservatorio Astronomico di Roma, Via Frascati 33, 00040 Monte Porzio Catone (RM), Italy
        \and
            INAF – Osservatorio Astronomico di Trieste, Via Tiepolo 11, 34143 Trieste, Italy
        \and 
            Fundación Galileo Galilei – INAF, Rambla José Ana Fernandez Pérez 7, 38712 Breña Baja, TF, Spain
        \and 
            Instituto de Astrofísica de Canarias (IAC), C/Vía Láctea s/n, E-38205, La Laguna (Tenerife), Canary Islands, Spain
        \and
            Departamento de Astrof\'{\i}sica, Univ. de La Laguna, Av. del Astrof\'{\i}sico Francisco S\'anchez s/n, E-38205 La Laguna (Tenerife), Canary Islands, Spain
        \and
            INAF - Osservatorio Astronomico di Padova, Vicolo dell'Osservatorio 5, IT-35122, Padova, Italy
        \and 
            INAF – Osservatorio Astronomico di Capodimonte, via Moiariello,
            16, 80131 Naples, Italy
        \and 
            Department of Physics, University of Rome “Tor Vergata”, Via della Ricerca Scientifica 1, 00133 Rome, Italy
        \and 
            Max Planck Institute for Astronomy, Königstuhl 17, 69117 Heidelberg, Germany
        \and 
            INAF – Osservatorio Astronomico di Cagliari, Via della Scienza 5,
            09047 Selargius, CA, Italy
        }

   \date{Received; Accepted}

\titlerunning{Spot modeling of V1298Tau using SpotCCF tool}
\authorrunning{Di Maio et al.}
 
\abstract
  {The intrinsic variability due to the magnetic activity of young active stars is one of the main challenges in detecting and characterising exoplanets.
  The stellar activity is responsible for jitter effects observed both in photometric and spectroscopic observations that could impact our planetary detection sensitivity.}
  {We present a method able to model the stellar photosphere and its surface inhomogeneities (starspots) in young/active and fast-rotating stars, based on the cross-correlation function (CCF) technique, to extract information about the spot configuration of the star.
  }
  {We developed \texttt{SpotCCF}, a tool able to model the deformation of the CCF profile due to the presence of multiple spots on the stellar surface. Within the Global Architecture of Planetary Systems (GAPS) Project at the Telescopio Nazionale Galileo, we analysed more than 300 spectra of the young planet-hosting star V1298 Tau provided by HARPS-N high-resolution spectrograph. By applying the \texttt{SpotCCF} model to the CCFs we extracted the spot configuration (latitude, longitude and projected filling factor) of this star, and also provided the new RVs time series of this target. 
  }
  {We find that the features identified in the CCF profiles of V1298 Tau are modulated by the stellar rotation, supporting our assumption that they are caused by starspots. The analysis suggests a differential rotation velocity of the star with lower rotation at higher latitudes. Also, we find that \texttt{SpotCCF} provides an improvement in RVs extraction with a significantly lower dispersion with respect to the commonly used pipelines, with consequent mitigation of the stellar activity contribution modulated with stellar rotation. A detection sensitivity test, by the direct injection of a planetary signal into the data, confirmed that the \texttt{SpotCCF} model improves the sensitivity and ability to recover planetary signals.
  }
   {Our method enables the modelling of the stellar photosphere, extracting the spot configuration of young/active and rapidly rotating stars. It also allows for the extraction of optimised RV time series, thereby enhancing our detection capabilities for new exoplanets and advancing our understanding of stellar activity.
   }

   \keywords{Stars: activity -- starspots -- Techniques: radial velocities -- Techniques: spectroscopic}

   \maketitle
%
\section{Introduction}
 Stellar activity in late-type main-sequence stars is the observable evidence of a hydromagnetic dynamo generating magnetic fields in their external convection zones. It manifests through a variety of phenomena that reach the stellar surface (e.g. chromospheric plages, starspots, heating of the chromosphere and corona, and impulsive flares) and can be indirectly observed with numerous activity tracers, including Ca II H\&K emission \citep[e.g.][]{Noyes1984ApJ...279..763N, Scandariato2017A&A...598A..28S, Maldonado2017A&A...598A..27M, DiMaio2020A&A...642A..53D},  X-ray flux \citep[e.g.][]{Wright2011ApJ...743...48W}, or photometric variability caused by spots and flares \citep[e.g.][]{Kron1947PASP...59..261K, Walkowicz2013ApJS..205...17W, Davenport2016ApJ...829...23D}. 

 Starspots, regions cooler than the surrounding photosphere, represent a manifestation of magnetic field lines going through the stellar photosphere and obstructing the convective welling up of hot plasma \citep{Schrijver1989ApJ...337..964S, Skumanich1975ApJ...200..747S, Solanki2006RPPh...69..563S, He2018ApJS..236....7H, Choudhuri2017SCPMA..60a9601C}. 
 In general, stellar activity leads to temperature inhomogeneities on the stellar surface and a quenching of the convective blueshifts of photospheric spectral lines.
 
 A significant fraction of young solar-type stars show very fast rotation rates, which play a key role in amplifying internal magnetic fields through the action of the dynamo mechanism. 
 However, in highly fast rotating stars, the level of activity saturates \citep[e.g.][]{Pizzolato2003A&A...397..147P, Vidotto2014MNRAS.441.2361V, See2019ApJ...876..118S}. Several theoretical attempts have been made to model this specific dynamo state \citep[e.g.][]{Kitchatinov2015RAA....15.1801K}; however, a conventional explanation for this effect is still lacking.

 In this context, spot size and distribution patterns on the stellar surface could serve as fundamental constraints for understanding the stellar dynamo mechanisms \citep{Strassmeier2009A&ARv..17..251S}. 
Additionally, many works have explored the possibility offered by star spots as tracers of stellar rotation and differential rotation \citep{Mosser2009A&A...506..245M, Walkowicz2013ApJS..205...17W, Lanza2016LNP...914...43L, Mancini2017MNRAS.465..843M}, which are essential components of theoretical models explaining the amplification mechanisms of the stellar dynamo \citep[e.g.][]{Spruit2002A&A...381..923S, Arlt2011MNRAS.412..107A}.  

Stellar activity has a significant influence on the stellar environment, and in turn, on the formation and early evolution of planets orbiting the star. A detailed understanding of stellar activity is a prerequisite for exoplanet detection through indirect methods, in particular with the radial velocity (RV) technique, which rely on the efficient mitigations of the activity noise. Indeed, stellar activity can pose challenges for exoplanet detection and characterization: the presence of spots can alter photospheric spectral lines, inducing radial velocity variations that can either mask planetary signals or produce spurious signals that mimic planet signatures.  
Planetary systems at young ages, when the activity level is strong enough to hamper planet detection, represent an interesting resource to understand the formation and migration process, as well as for the study of the physical evolution of the planets themselves and the planet's photoevaporation. Although the high level of stellar activity makes more challenging the search for new exoplanet candidates, it is important to monitor and study young and intermediate-age stars to search for planets in formation or at the early stage of their evolution within the timescale of migration. 

In this context, modelling the activity of young stars is crucial to increase our knowledge about the early evolution of stars and their planets and also our ability to disentangle stellar and planetary signals, allowing the detection of new exoplanet candidates. 
Many techniques for observing and modelling stellar activity have been developed. Traditionally, star-spots are detected using modelling techniques, e.g. light-curve inversion \citep[e.g.][]{Budding1977Ap&SS..48..207B, Savanov2008AN....329..364S, Lanza2010A&A...520A..53L, Scandariato2017A&A...606A.134S}, Doppler imaging \citep[e.g.][]{CollierCameron1994MNRAS.269..814C, Strassmeier2000A&A...360.1019S, Strassmeier2002AN....323..309S}, Zeeman-Doppler imaging \citep[e.g.][]{Donati1998ASPC..154.1966D, Donati2003MNRAS.345.1145D, Strassmeier2023A&A...674A.118S}. 

Stellar activity can be also probed by measuring time-dependent variations in the shape of the cross-correlation function (CCF). The CCF is computed from the correlation between the measured spectrum with a weighted binary mask and is constructed by shifting the mask as a function of the Doppler velocity. The binary mask excludes the strongest spectral lines such as H$\alpha$ or Ca II H\&K or the NA doublet that can have a chromospheric reversal in their cores.

The resulting CCF is a function describing a flux-weighted mean profile of the stellar absorption lines transmitted by the mask. The minimum of the CCF as a function of the Doppler offset is the RV measurement. The shape of the CCF is usually fitted to find the stellar radial velocity using a Gaussian profile \citep{Fellgett1955AcOpt...2....9F, Baranne1996, Pepe2002}. 

Many authors tried to correct for stellar activity when extracting and analysing the radial velocity using several approaches: de-correlating the RV measurements against activity indicators, such as log R$^{'}_{\mathrm{HK}}$ or line shape indicators \citep[e.g][]{Saar2000ApJ...534L.105S, Meunier2010, Meunier2017, Lanza2018, Maldonado2019A&A...627A.118M}; modelling stellar activity in RVs with Gaussian process, moving average, or kernel regression technique \citep[e.g.][]{Haywood2014MNRAS.443.2517H, Rajpaul2015MNRAS.452.2269R, Lanza2018, Damasso2023A&A...672A.126D}; using a line-by-line analysis \citep[e.g.][]{Dumusque2018, Cretignier2020A&A...633A..76C}, studying spectral line profile variability \citep[e.g.][]{Zhao2019, Meunier2020}. 
In this paper, we present \texttt{SpotCCF}, a new method for spot modelling and RV extraction based on the CCF fitting. \texttt{SpotCCF} is developed focusing on active stars with a significant rotational broadening of the order of a few tens of km s$^{-1}$. In the context of the Global Architecture of Planetary System (GAPS) program \citep{Covino2013A&A...554A..28C, Carleo2020A&A...638A...5C}, we applied our method to the HARPS-N observations of the young planet-hosting star V1298 Tau.
The reduction of the high-resolution HARPS-N observations and the extraction of the radial velocities are typically performed by using the Data Reduction Software (DRS) pipeline \citep{Pepe2002, Dumusque2021A&A...648A.103D} and in specific cases also by HARPS-TERRA (Template-Enhanced Radial velocity Red-analysis Application) pipeline \citep{TERRA2012}. 
In young and highly-rotating stars, the spectral lines are enlarged due to the rotational broadening which dominates other broadening effects. For this reason, the CCF profile is not accurately described by the Gaussian fit, as assumed by the DRS pipeline, but rather by a rotational profile.
Moreover, since the CCF profile gathers information from each individual spectral line selected in the mask, average changes in the individual line profiles due to stellar activity produce changes in the CCF shape. The net result is that the average CCF profile changes from one observation to another (depending on the instantaneous spot configuration) making not accurate the usage of a reference template as done in TERRA. This circumstance motivated the implementation of an alternative RV extraction pipeline.

This paper is organised as follows. We describe the target and the observations in Sect. \ref{sec:V1298Tau}. We detail our model (\texttt{SpotCCF}) in Sect. \ref{sec:model}, and describe the correction of the CCF profiles and the RVs extraction in Sect. \ref{sec:ccfcorrection} and Sect. \ref{sec:extraction}, respectively. Section \ref{sec:analysis} presents the analysis of the RVs time series and the comparison with the TERRA dataset. In Sect. \ref{sec:spotanalysis} we characterised the spots using the parameter obtained from the fit, while a test of the presence of differential rotation follows in Sect. \ref{sec:diffrot}. 
 In Sect. \ref{sec:injection} we tested the detection sensitivity by direct injection of a planetary signal into the data.  Our conclusions follow in Sect. \ref{sec:summary}.

 \section{V1298 Tau}\label{sec:V1298Tau}
 V1298 Tau is a young solar-mass K1 star, relatively bright with a visual magnitude of 10.1.  Its estimated effective temperature is 5050 $\pm$ 100 K, with solar metallicity, and a luminosity of 0.954 $\pm$ 0.040 $L_\odot$ \citep{SuarezMascareno2022NatAs...6..232S}. The logarithmic surface gravity is  4.246 $\pm$ 0.034 dex \citep{David2019AJ....158...79D}. It is located at a distance of 108.6 $\pm$ 0.7 pc in the Taurus region and belongs to the Group 29 stellar association \citep{Oh2017AJ....153..257O}. The main stellar parameters of V1298 Tau are listed in Table  \ref{table:V1298Tau}. 
\begin{table}[h!]\small
	\caption{V1298 Tau main parameters}   
	\label{table:V1298Tau} 
	\centering             
	\begin{tabular}{cc}   
		\toprule[0.05cm]
		\toprule
		\multicolumn{2}{c}{V1298 Tau} \\
		\midrule            
		\smallskip
		Spectral type & K1 \\ 
		\enskip
		$\mathrm{M_{\star}}\mathrm{(M_{\odot})}$ \ \tablefootmark{(a)} &$1.170 \pm 0.060$\\
		\enskip
		$\mathrm{R_{\star}}\mathrm{(R_{\odot})}$ \ \tablefootmark{(a)} &$ 1.278 \pm 0.070$ \\
		\enskip
		$\log g$ (dex) \ \tablefootmark{(b)} & $4.246 \pm 0.034$ \\
		\enskip
		Age (Myr)\tablefootmark{(b)} & $23 \pm 4 $ \\
		\enskip
		$\mathrm{T_{\rm eff}}$ (K) \ \tablefootmark{(a)} & $5050 \pm 100$ \\
		\enskip
		distance (pc)\ \tablefootmark{(c)} & $\sim 108.5 \pm 0.7$ \\
		\enskip
		$v$  sin $i$ $\mathrm{(km \ s^{-1})}$\tablefootmark{(a)}& $23.8 \pm 0.5$ \\
		\enskip
		$\mathrm{P_{\rm phot}}$ (d)\ \tablefootmark{(b)} & $  2.865\pm 0.012$ \\
		\bottomrule[0.05cm]                
	\end{tabular}
	\tablefoot{\tablefoottext{a}{\citet{SuarezMascareno2022NatAs...6..232S}} \tablefoottext{b}{\citet{David2019AJ....158...79D}} \tablefoottext{c}{\citet{Gaia2021A&A...649A...1G}}.}
\end{table}

V1298 Tau was observed in 2015 by NASA's K2 mission \citep{Howell2014PASP..126..398H}, which led to the discovery of four transiting planets, all with sizes ranging from Neptune to Jupiter \citep{David2019AJ....158...79D, David2019ApJ...885L..12D}. The three inner planets have well-constrained orbital periods of 24.1396 $\pm$ 0.0018, 8.24958 $\pm$ 0.00072 and 12.4032 $\pm$ 0.0015 days, and radii of  0.916$^{+0.052}_{-0.047}$, 0.499$^{+0.032}_{-0.029}$ and 0.572$^{+0.040}_{-0.035}$ R$_{\rm J}$, respectively. The outer planet, e, is the most challenging planet of the system, with a more uncertain orbital period (40-120 days) and a radius of 0.780$^{+0.075}_{-0.064}$ R$_{\rm J}$, detected with a single transit. \citet{Feinstein2022ApJ...925L...2F}, by using a new TESS observation, found the most likely period of V1298 Tau e to be 44.17 days, and a radius 9.94 $\pm$ 0.39 R$_{\rm J}$, $\sim3\sigma$ larger than what was found in the original K2 data \citep{David2019ApJ...885L..12D}. 

Given the youth of the system and its potential to provide insights into the initial conditions of close-in planetary systems \citep[e.g.,][]{Owen2020MNRAS.498.5030O, Poppenhaeger2021MNRAS.500.4560P}, extensive follow-up observations have been conducted on V1298 Tau. These efforts include attempts to constrain planet masses through radial velocities \citep{Beichman2019RNAAS...3...89B, SuarezMascareno2022NatAs...6..232S, Finociety2023MNRAS.526.4627F}, measurement of the spin-orbit alignments of planet c \citep{Feinstein2021AJ....162..213F} and planet b \citep{Gaidos2022MNRAS.509.2969G, Johnson2022AJ....163..247J}, measurement or constraint of atmospheric mass-loss rates for the innermost planets \citep{Schlawin2021RNAAS...5..195S, Vissapragada2021AJ....162..222V, Maggio2022ApJ...925..172M}, and an approved program to study planetary atmospheres using the James Webb Space Telescope \citep{Desert2021jwst.prop.2149D}.

V1298 Tau is a highly active star that exhibits significant radial velocity variations, principally caused by stellar activity. It is a star with a high projected equatorial rotational velocity, $v \sin i$ of 23 $\pm$ 2 km s$^{-1}$ \citep{David2019AJ....158...79D}, and displays a broadened CCF profile.  This derived rotation velocity is consistent with the value of 23.8 $\pm$ 0.5 km s$^{-1}$ reported by \citet{SuarezMascareno2022NatAs...6..232S}. 

In this work, we analysed 311 spectra of V1298 Tau, obtained with the HARPS-N spectrograph \citep{Cosentino2014SPIE.9147E..8CC} within the GAPS observing program. 
HARPS-N is fibre-fed and has a resolving power of $R$ $\sim$ 115 000 and covers a spectral range of 3830-6930 $\AA$.
The spectra were collected, with the acquisition mode involving fibre A on the target and fibre B on the sky,  between March 2019 and March 2023, with a typical exposure time of 1200 s, and a median S/N of 60 measured at a wavelength of 5500 $\AA$.
When permitted by the visibility window of the target, we collected two spectra on the same night, separated by at least two hours. The spectra were reduced with the K5 mask using the YABI platform \citep{YABI} hosted at the IA2 Data Center\footnote{https://www.ia2.inaf.it}. 
In addition, due to the high rotation rate of the star, we increased the width of the half-window for the computation of the CCF from the default value of 20 up to 100 km s$^{-1}$ to clearly cover the continuum around the wings of the CCF profile.

The typical RV dispersion ranges between 890 ms$^{-1}$ and 290 ms$^{-1}$ for DRS and TERRA, respectively. Some of these observations were included in the dataset analysed by \citet{SuarezMascareno2022NatAs...6..232S}.
More details on the HARPS-N time series are provided in Section \ref{sec:analysis}.
\section{Modeling of CCF in the presence of spots}\label{sec:model}
 The aim of this work is to develop a model (\texttt{SpotCCF}) representing the CCF profile of active stars, with $v \sin i$ of tens of km s$^{-1}$ in which the CCF is strongly broadened. In these cases, the spectral lines are enlarged due to the rotational broadening, whose characteristic dominates when rotation is greater than other stellar broadening effects. For this reason, the CCF is not accurately described by the Gaussian fit, as employed by the HARPS/HARPS-N DRS pipeline, but rather by a  rotation profile. 

 Assuming the star to be spherical and rotating as a rigid body, a synthetic spectrum can be obtained through the convolution of the spectrum of a non-rotating star with the rotational profile $G(\Delta \lambda)$ described by Gray’s equation \citep{Gray2018}.

 \begin{equation}
    G(\Delta\lambda) = G(v_{\rm z}) = \begin{cases}\dfrac{1}{v_{\rm L}} \dfrac{\int_{-y_{\rm 1}}^{+y_{\rm 1}}I_{\rm c} dy/R}{\oint I_{\rm c}\cos\theta d\omega} & \text{if } \left | v_{\rm z} \right | \leq v_{\rm L} \text{ or } \left | \Delta\lambda \right | \leq \Delta\lambda_{\rm L} \\
    \qquad \qquad 0  & \text{if } \left | v_{\rm z} \right | > v_{\rm L} \text{ or } \left | \Delta\lambda \right | > \Delta\lambda_{\rm L} \end{cases}
    \label{eq:graysint}
 \end{equation}
 with $v_{\rm L} = v \sin i$ where $v$ is the equatorial rotation velocity, and $i$ is the angle of inclination of the star spin axis with respect to line-of-sight.  $v_{\rm z}$ represents the $\Delta\lambda$ shift from the line centre expressed in kilometres per second. The continuum intensity is denoted as $I_{\rm c}$, while $R$ represents the stellar radius.  The stellar disc can be divided into strips parallel to the projection of the spin axis on the disc itself, and the  Doppler shift is constant along each strip, with the highest shift at the limbs with value $v_{\rm L} = v \sin i$. The integration limits, $y_{\rm 1}$, represent the extremes of the projected stellar disc in the orthogonal plane where the projections are defined, considering that the origin is in the star centre. We evaluate $G(\Delta\lambda)$  assuming the limb darkening law proposed by \citet{Claret2000A&A...363.1081C} (Eq. \ref{eq:limbdarkening}), which is the most precise analytical representation of the limb darkening to date \citep{Howarth2011MNRAS.413.1515H, Morello2017AJ....154..111M}. 
 \begin{equation}
	\label{eq:limbdarkening}
	\dfrac{I_{\rm \lambda(\mu)}}{I_{\rm \lambda(1)}} = 1 - \sum_{n=1}^{4} a_{\rm n,\lambda} (1-\mu^{n/2}),	
\end{equation}
where $\lambda$ indicates a specific spectral bin/passband; $\mu = \cos \theta$, being $\theta$ the angle between the stellar surface normal and the line of sight.  The stellar intensity is denoted with $I_{\rm \lambda(\mu)}$, while $I_{\rm \lambda(1)}$ is the intensity at the centre of the disc ($\mu$ = 1). The limb darkening coefficients, $a_{\rm n,\lambda}$, are derived using the database \texttt{PHOENIX\textunderscore2012\textunderscore13} \citep{Claret2012A&A...546A..14C,Claret2013A&A...552A..16C} of the \texttt{ExoTETHyS} package \citep{Morello2020AJ....159...75M}. 
\begin{figure}[h!]
    \centering
    \includegraphics[width=0.83\hsize]{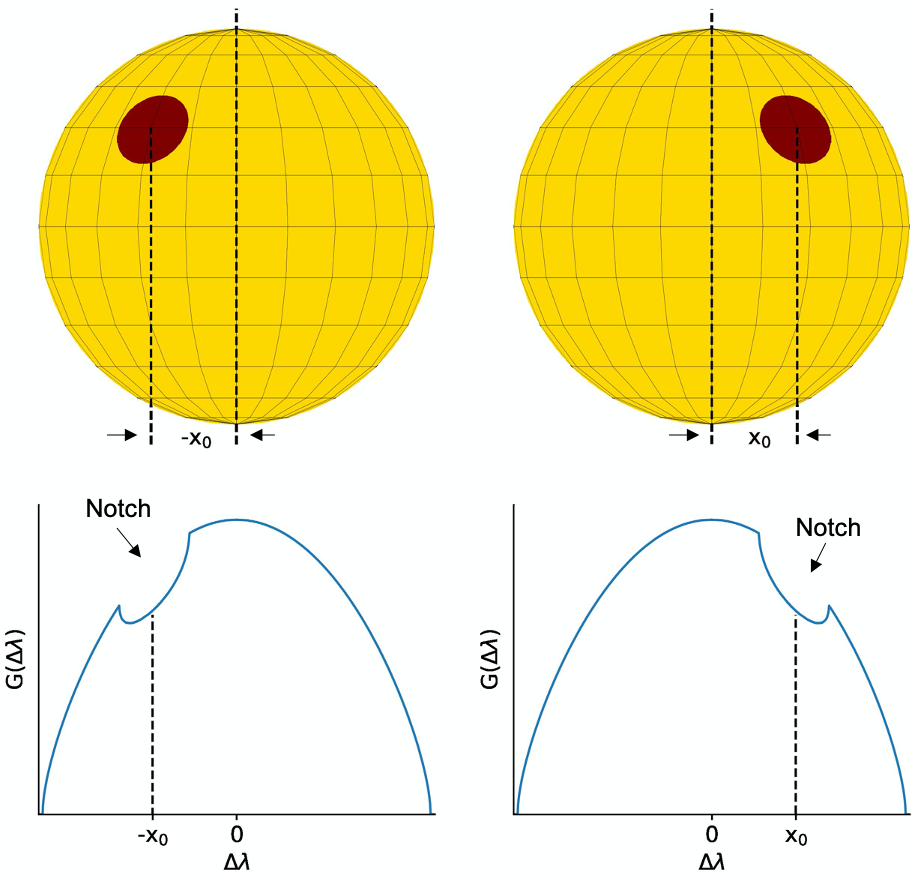}
    \caption{(Upper panel) Spot centered at (-x$_{\rm 0}$,y$_{\rm 0}$) on the projected stellar disc. The rotation of the star brings the spot centre to the new position on the disc (x$_{\rm 0}$,y$_{\rm 0}$). (Bottom panel) Rotation profile modified by the presence of a spot. The Doppler-shift distribution exhibits a notch or dip at the spot's position. As rotation carries the spot across the stellar disc, the position of the notch moves across the Doppler-shift distribution and its strength changes with the projected area of the spot and for the effect of the limb darkening. Figure adapted from \citet{Gray2018}.}
    \label{fig:bumpgraycartoon}%
    \end{figure}
 
Stars with high $v \sin i$ show higher stellar activity levels than other stars. For this reason, to take into account the activity level of the star, the presence of one or more spots on the stellar disc was included in the model. 
 
 In Fig. \ref{fig:bumpgraycartoon} we sketch the correspondence between position within the Doppler-shift distribution and longitude on the star for an equator-on line of sight. Surface features that alter the amount of light coming from specific longitude bands will introduce structures into the rotational profile. In the simplest case, a dark spot reduces the contribution to the light in one of the strips, as shown in Fig. \ref{fig:bumpgraycartoon}.
 This produces a dip or notch in the rotation function $G(\Delta\lambda)$ at the Doppler shift corresponding to that strip. The bump appears on the short-wavelength side of the profile as the spot becomes visible on the approaching limb. It migrates across the profile, growing stronger as the spot becomes more nearly face-on, reaching its largest size as the spot crosses the star's meridian. As the bump continues on the receding side of the disc, the same pattern is played in reverse, the bump fading in size and moving to its maximum positive Doppler shifts. More generally, the Doppler shift for a spot of latitude $l$ and longitude $L$ is given by 
\begin{equation}
	\Delta \lambda = v \sin i \cos l \sin L \ 
\end{equation}

 To take into account the presence of the spot we adapted Gray’s formula (Eq. \ref{eq:graysint}) changing the integration limits and computing a rotational profile only across the portion of the disc covered by the spot.
 In this way, we obtained the analytical expression of Gray’s formula for the spot, $G_{\rm spot}(\Delta \lambda)$.

 By subtracting the Gray rotational profile of the spot, $G_{\rm spot}(\Delta\lambda)$, from the rotation profile calculated on the entire disc $G(\Delta\lambda)$, we obtain the final rotation profile $G_{\rm f}(\Delta\lambda)$, which takes into account the presence of the spot.
 \begin{equation}
    G_{\rm f}(\Delta\lambda) = norm * [G(\Delta\lambda) - G_{\rm spot}(\Delta\lambda)],
 \end{equation}
 where $norm$ is a normalisation parameter.

However, the analytical calculation of this function is computationally demanding, which is why a numerical model was developed. We constructed a photospheric stellar model, where the star is treated as a spherical object that is mapped onto a Cartesian coordinate system, with a grid of 999 pixels in the range [-1;1] in unit of R$_{\star}$, and each spot is represented as a spherical cap with a radius matching that of the spot. The stellar flux contribution is calculated for each pixel on the stellar map, with no contribution from the pixels representing the spots (black spots, $T \approx$ 0). Finally, the calculated flux contribution is integrated across each strip of the stellar disc.

 To model the CCF, we need to establish the latitude and the longitude of the centre of the spot, as well as the filling factor, to identify where the $G_{\rm spot}(\Delta\lambda)$ should be calculated, or in the case of the stellar map, the pixels of the grid representing the spot. 

 If we use a coordinate system where the $z$-axis is along the line of sight, $x$ and $y$ identify the orthogonal plane where the projections are defined (the plane of the sky) and the origin is the star centre, the projected spot centre is in the position ($x_{\rm 0}$,$y_{\rm 0}$)
 \begin{equation}
    \begin{cases}
        x_{\rm 0} = \sin(l)\sin(i)-\cos(l)\cos(L)\cos(i)\\
        y_{\rm 0} = \cos(l)\sin(L),
    \end{cases}
 \end{equation}
where $l$ and $L$ are the latitude and longitude of the spot, respectively, and $i$ is the inclination of the stellar spin to the line of sight $z$ (see a detailed derivation in Appendix \ref{appendice:modello}). 
Since we are assuming a spot with a circular shape, its projection on the stellar disc will be an ellipse with semi-axes $a = R_{\rm spot}$ and $b = R_{\rm spot} \cos(l) \sin(L)$, where $R_{\rm spot}$ is the radius of the spot in stellar radii units. The spot-filling factor is $ff = R_{\rm spot}^2$ while the filling factor projected on the stellar disc is the area covered by the ellipse, $ff_{\rm p} = ab = R_{\rm spot}^2\sqrt{1-r^2}$, where $r=\sqrt{x_{\rm 0}^2+y_{\rm 0}^2}$ is the projected distance of the spot from the star centre. 
 We used the relation between the rotational period of the star and the amplitude of the rotational modulation obtained by \citet{Messina2001A&A...366..215M} to put some constraints on the filling factor of the spot, considering $P_{\rm rot} = \dfrac{2\pi R_{{\star}}}{v}$.

The numerical integration does not take into account the wings of the CCF profile because in those points the function is not defined. For this reason, the $G_{\rm f}(\Delta \lambda)$ has been convolved with a Lorentzian function, L($\Delta \lambda$), that takes into account the wings in the CCF profile, given by
\begin{equation}
	L(\Delta \lambda, \gamma) = \dfrac{1}{\pi\gamma}\left[\dfrac{\gamma^2}{\Delta \lambda^2+\gamma^2}\right],
\end{equation} 
where $\gamma$ is the scale parameter which specifies the half-width at half-maximum.

 Finally, the model used to fit the CCF profile was described by 
\begin{equation}
	Model = G_{\rm f}(\Delta \lambda) \ast L(\Delta \lambda).
\end{equation}

\section{CCFs correction}\label{sec:ccfcorrection}
When the "objAB" acquisition mode is employed for HARPS-N, the fibre B of the spectrograph is used to acquire 
\begin{figure}[h!]
	\centering \includegraphics[scale=0.135]{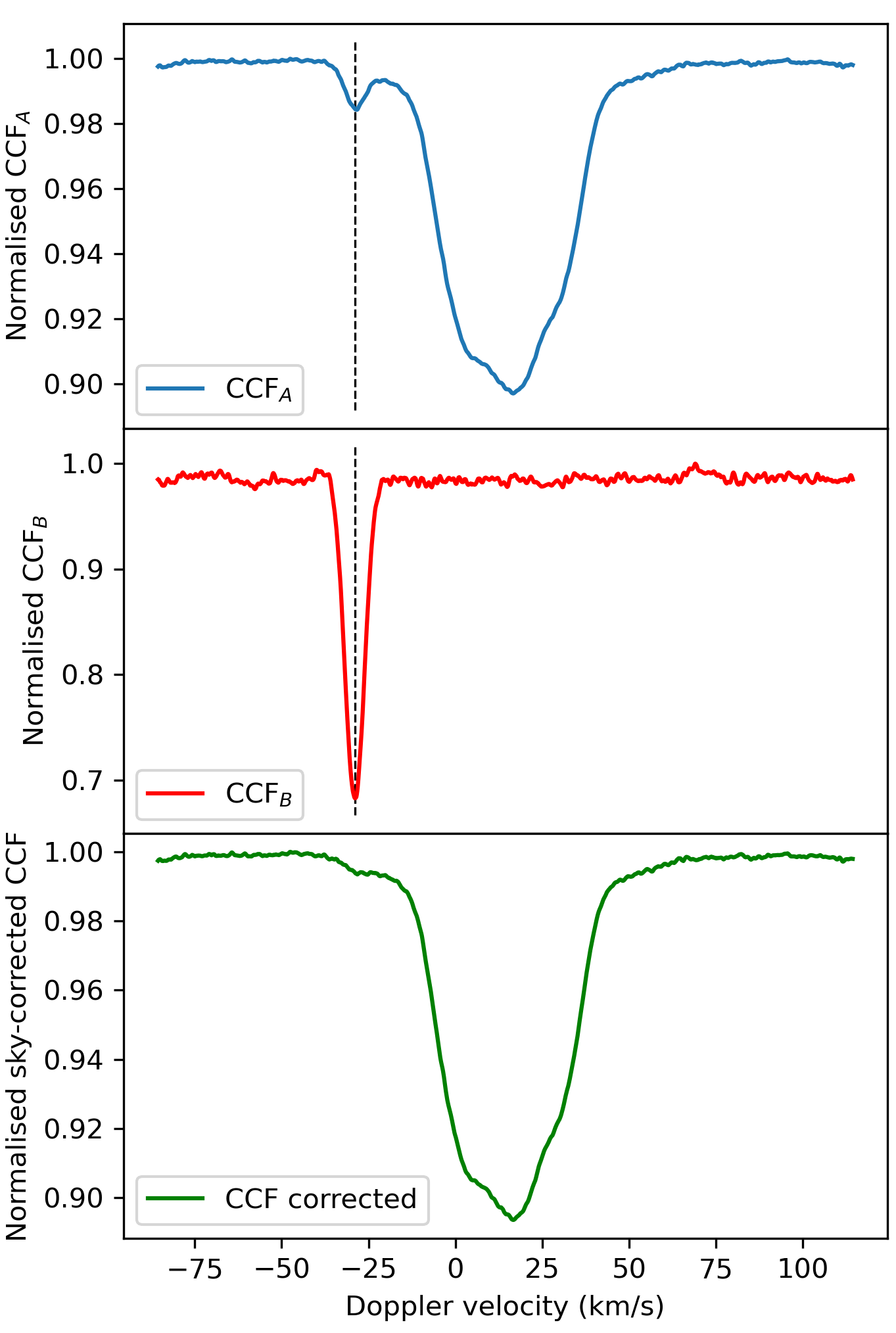}
	\caption{Example of CCF profile distorted by the effect of the moon illumination. (Top) CCF profile of the fibre A. (Middle) CCF profile of fibre B, CCF$_{\rm B}$. The black dashed vertical lines highlight the deformations in both profiles. (Bottom) Sky-corrected CCF.}
    \label{fig:correzioneCCFB}
\end{figure}
the sky spectrum, in order to obtain an optimal subtraction of the detector noise and background.
The DRS pipeline computes also the CCF of the sky spectrum obtained using the fibre B, referred to as CCF$_{\rm B}$. 

Some of the V1298 Tau CCF profiles that were analysed exhibited anomalous deformations that were also present in the CCF$_{\rm B}$ (see an example in Figure \ref{fig:correzioneCCFB}). 
These types of deformations could appear in the wings or even the core of the CCF profile, leading to incorrect radial velocity estimations and/or introducing spurious signals. As previously reported by \citet{Malavolta2017AJ....153..224M}, these deformations may be caused by moon illumination. To mitigate these deformations in all the CCF profiles, we followed the procedure described in \citet{Malavolta2017AJ....153..224M}. Firstly, we recalculated the CCF$_{\rm B}$ using the same flux correction coefficients as those used for the target CCF$_{\rm A}$ during the specific acquisition. Next, the CCF$_{\rm B}$ was subtracted from the corresponding CCF$_{\rm A}$. The outcome of this process is the sky-corrected CCF profile (lower panel of Figure \ref{fig:correzioneCCFB}).

\section{Extraction of spot parameters and radial velocities}\label{sec:extraction}
\begin{figure}[htp!]
	\centering
	\includegraphics[width=0.9\hsize]{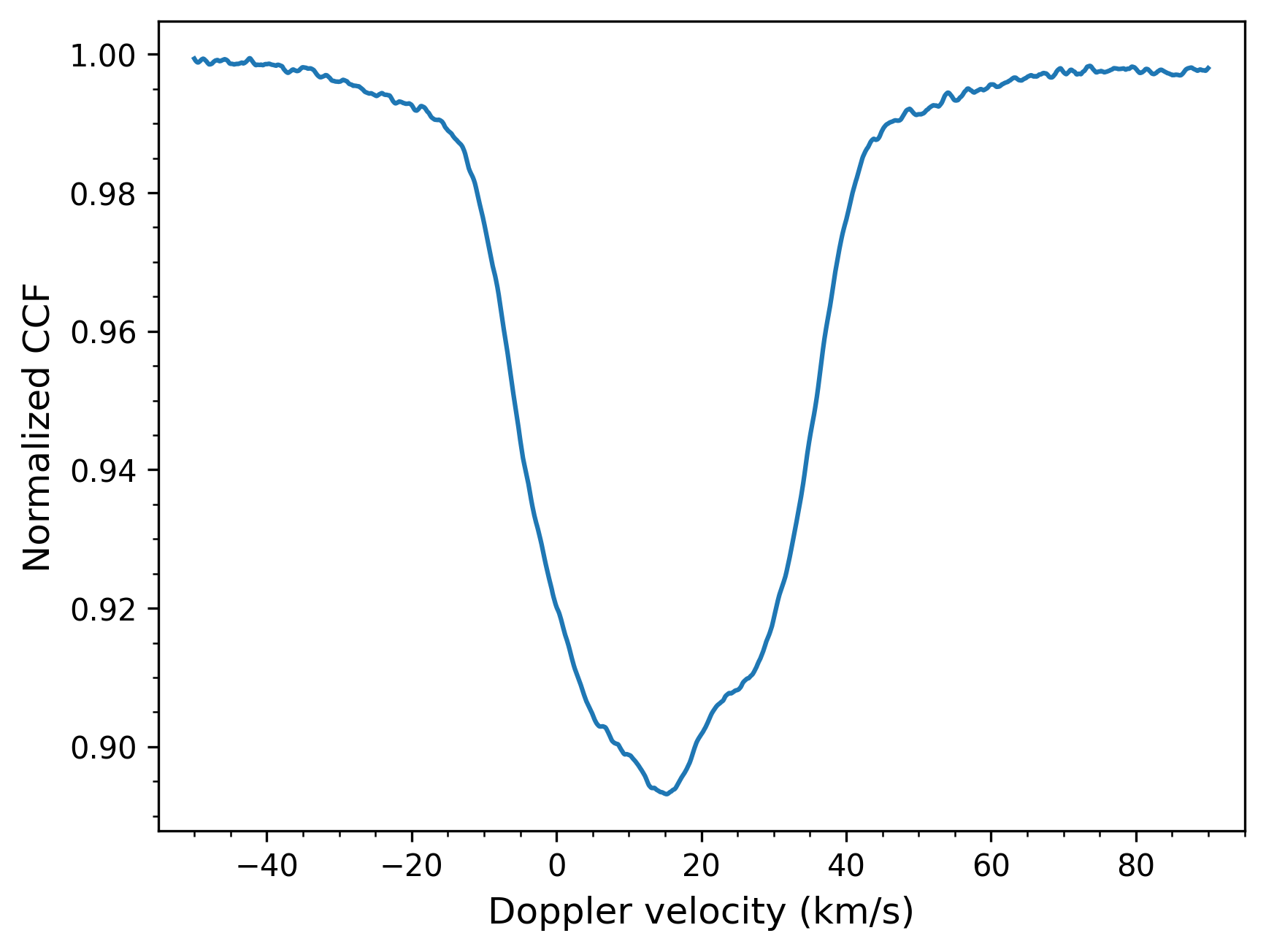}
	\caption[Example of a normalised CCF profile of V1298 Tau]{Example of a normalised CCF profile of V1298 Tau. The deformed profile is due to the presence of one or multiple spots.}
	\label{fig:exampleCCF}%
\end{figure}

The CCF profiles of V1298 Tau, corrected for sky effects, exhibit multiple deformations as shown in Figure \ref{fig:exampleCCF}.

We applied \texttt{SpotCCF} to the CCF profiles of V1298 Tau in two steps. In the first step, we used a model that accounted for the presence of one black spot on the stellar disc, which we refer to as the "One-spot model". We explored the full (hyper-)parameter space using the Monte Carlo (MC) nested sampler and Bayesian inference tool \texttt{MultiNest} v3.10 \citep{Feroz2019OJAp....2E..10F}, with the \texttt{pyMutliNest} wrapper \citep{Buchner2016ascl.soft06005B}. The priors used in all the analyses described below are summarised in Table \ref{tab:priorsSPOTS}. The MC sampler was set up to run with 5120 live points for both One-spot and Two-spots (see below) models and with a sampling efficiency of 0.5 for all cases considered in this study. The log-likelihood function to be minimised is expressed as:
 \begin{equation}
	\ln p(y_{\rm n},t_{\rm n},\theta) = -\dfrac{1}{2}\sum_{\rm n=1}^{N}\dfrac{[y_{\rm n}-f_{\rm \theta}(t_{\rm n})]^2}{\sigma^2_{\rm j}} -\dfrac{1}{2} \ln [2\pi (\sigma^2_{\rm j})],
\end{equation}
where $y_{\rm n}$ and $t_{\rm n}$ are the values of the CCF profile and the Doppler velocities, respectively; $\theta$ is the array of model parameters, $f_{\rm \theta}$($t$) is the model function, $N$ are the number of CCF points, and $\sigma_{\rm j}$ is the white noise term (jitter). 

In general, there are four parameters for the star ($v \sin{i}$, RV, $\gamma$, $i$),  three parameters for each spot (latitude, longitude, R$_{\rm spot}$), a normalisation parameter, and the jitter $\sigma_{\rm j}$. 

\begin{table}[h]\scriptsize
	\renewcommand*{\arraystretch}{1.3}
	\begin{center}
		\caption{Priors parameters of the model used for RVs extraction of V1298 Tau.}
		\label{tab:priorsSPOTS}
		\begin{tabular}{lcl}
			\toprule[0.05cm]
			Parameter    &       Prior     &       Description       \\
			\toprule[0.05cm]
			norm & $\mathcal{L}\mathcal{U}(10^{-4},10^2)$ & Normalisation parameter\\
			\midrule[0.01cm]
			\textit{Stellar parameters} & & \\
			RV (km s$^{-1}$) & $\mathcal{U}(10,20)$ & Centroid of CCF profile \\
			v $\sin{i}$ (km s$^{-1}$) & 24.74 (fixed) & Width of CCF profile\\
			$\gamma$ (km s$^{-1}$) & $\mathcal{U}(10^{-2},10^{2})$ & Lorentzian parameter \\
			$i$ (rad) & $\pi/2$ (fixed) & Inclinational angle \\
			\midrule[0.01cm]
			\textit{Spot parameters} & & \\
			latitude (rad) & $\mathcal{U}\biggl(0,\dfrac{\pi}{2}\biggr)$ & \\
			longitude (rad) & $\mathcal{U}(-\pi,\pi)$ & \\
			$R_{\rm spot}$ & $\mathcal{L}\mathcal{U}(10^{-2},0.5)$ & \\
			\bottomrule[0.05cm]
		\end{tabular}
		\tablefoot{The prior labels of $\mathcal{N}$, $\mathcal{U}$ and $\mathcal{LU}$ represent normal, uniform and loguniform distributions, respectively. Longitude and latitude are uniformly distributed in the interval.}
	\end{center}
\end{table}

\begin{figure}[h!]
	\centering
         {\subfloat[\label{fig:Onespotmodel}]{\includegraphics[width=0.85\hsize]{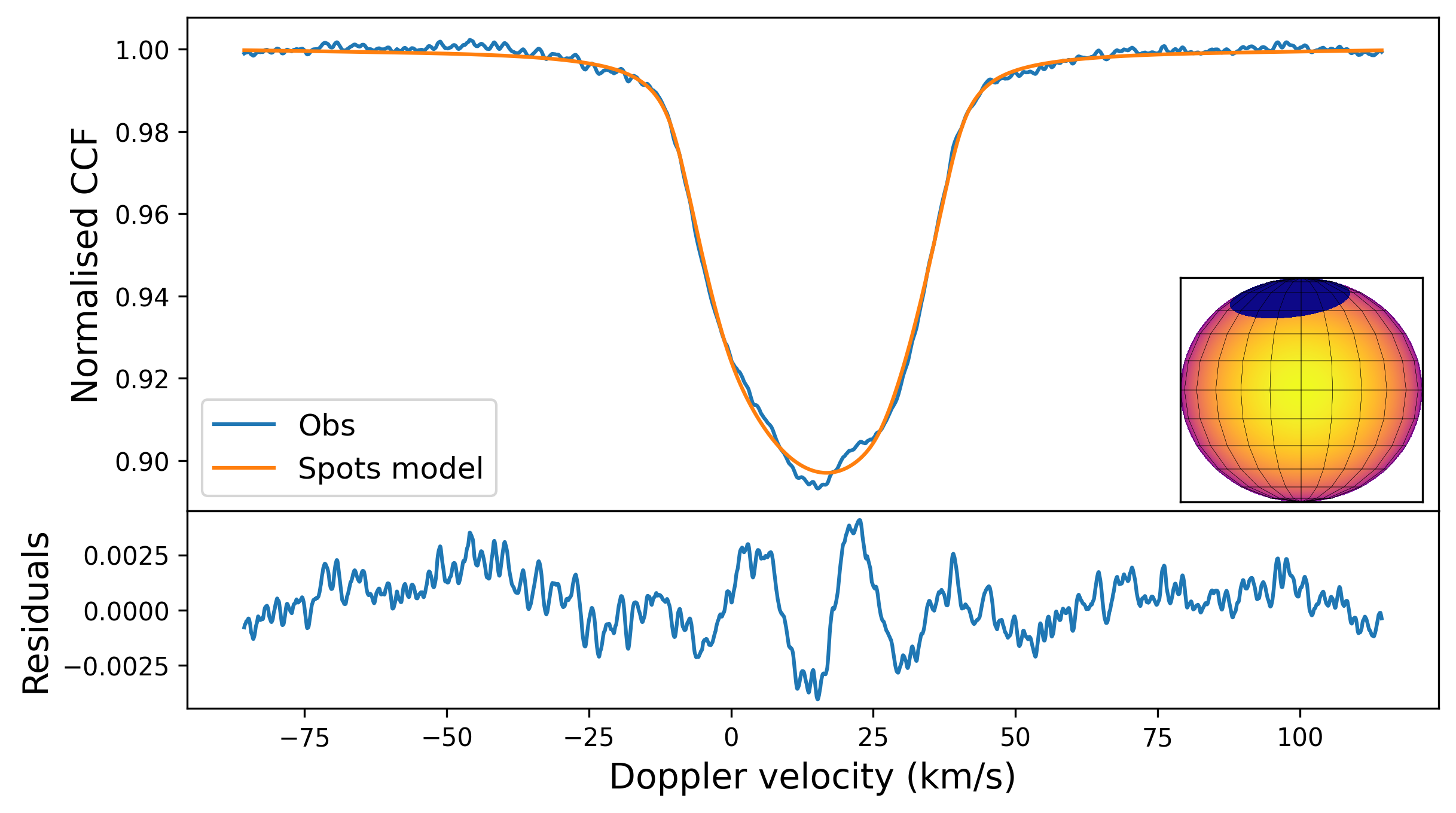}} \\
    \subfloat[\label{fig:Twospotsmodel}]{\includegraphics[width=0.85\hsize]{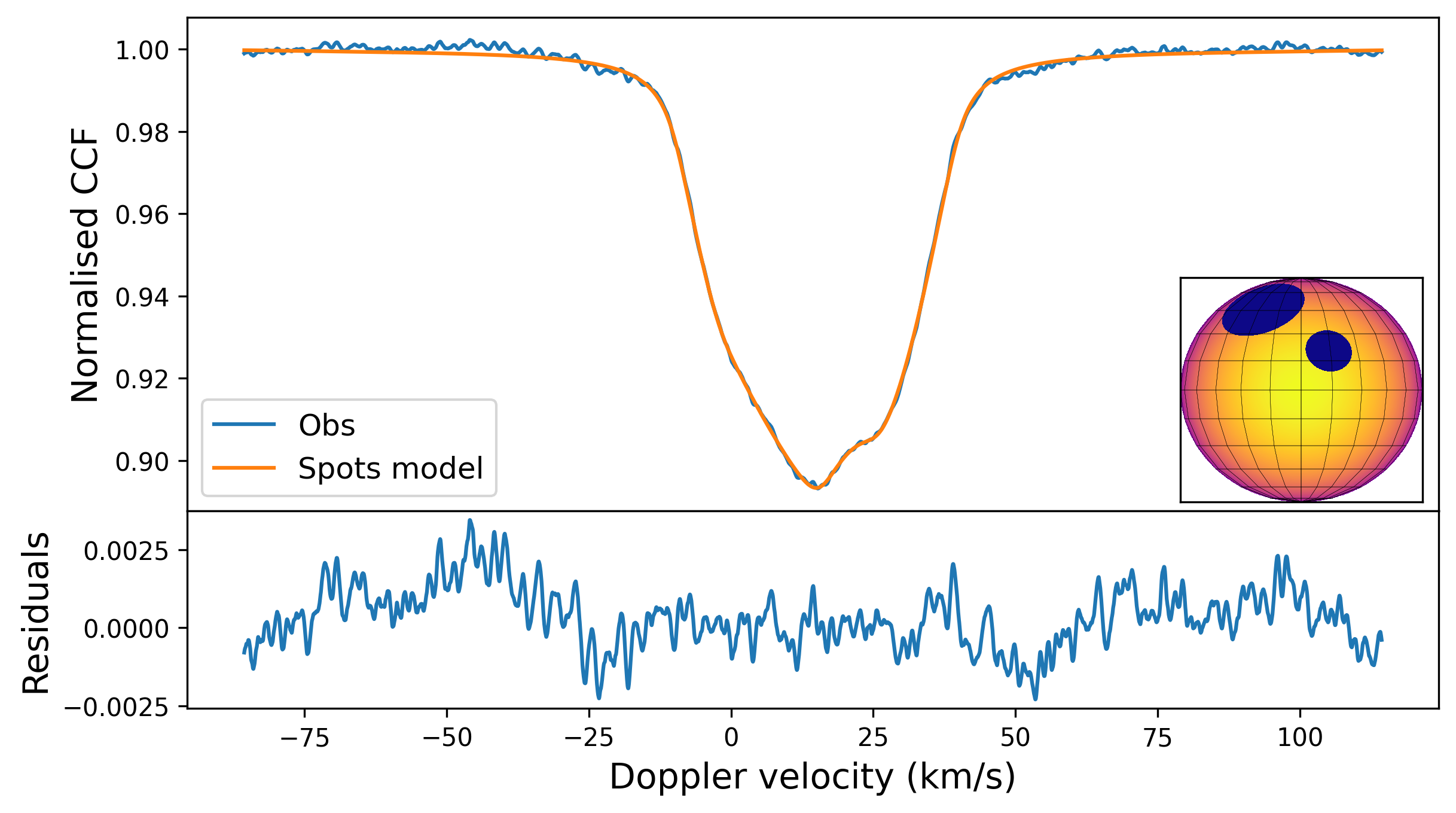}}
    }	\caption[Examples of CCF profiles of V1298 Tau fitted with "One-spot model" and "Two-spots model" and the corresponding spot configuration]{Examples of CCF profiles of V1298 Tau fitted with "One-spot model" (a) and "Two-spots model" (b) and the corresponding residuals (in bottom panels). The inset of each plot shows the location of the spots on the stellar disc, as defined by the fit of both models; the grid indicates longitudes and latitudes from -90 to 90 degrees with 15-degree intervals.}
	\label{fig:exampleCCFmodel}
\end{figure}
\begin{figure*}[ht!]
	\centering
	\includegraphics[width=0.9\hsize]{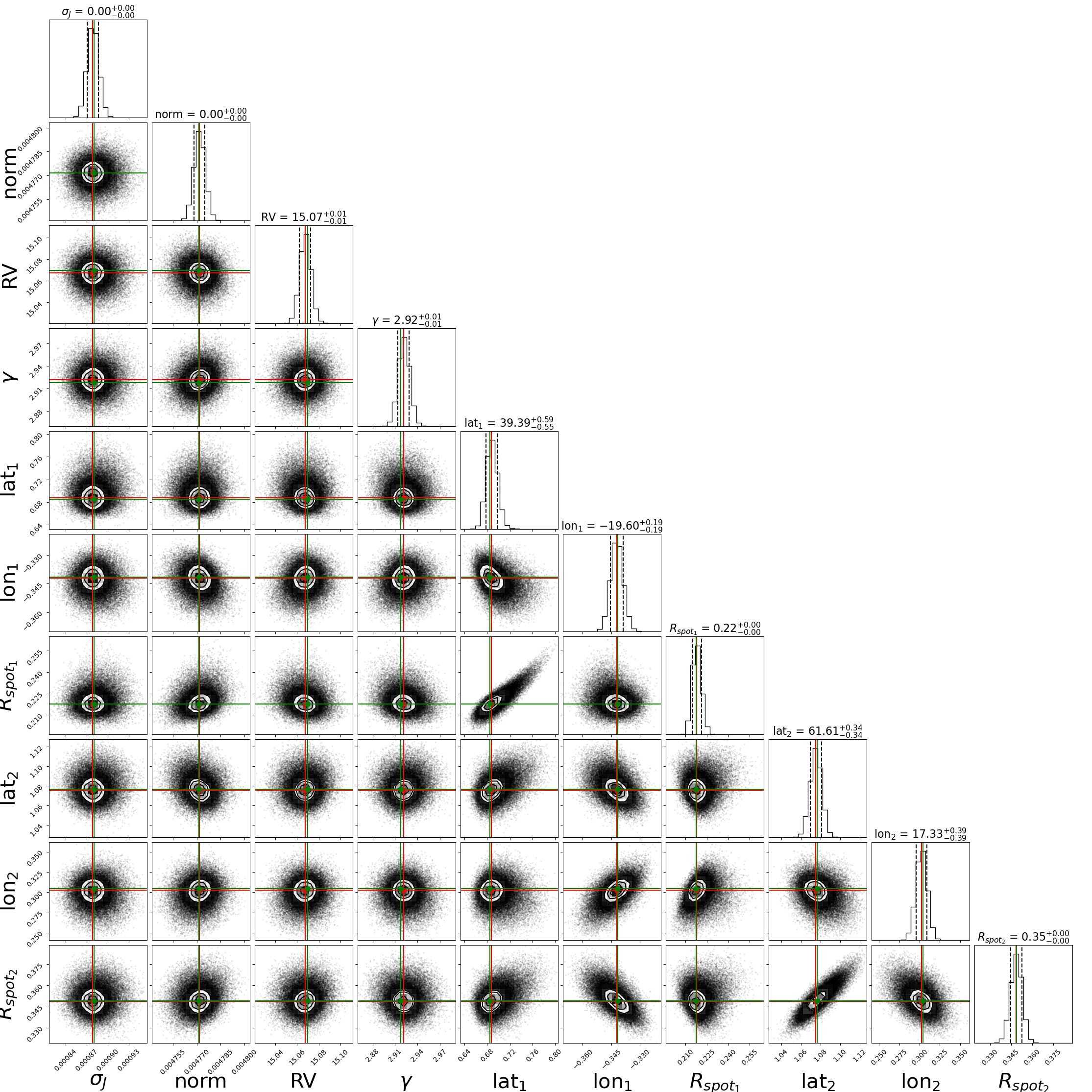}
	\caption{Example of corner plot of the best-fit parameters obtained from the fitting of a CCF profile of V1298 Tau with the "Two-spots model". The red and green lines mark the median and the maximum-a-posteriori values, respectively, while the dashed black lines are the 16$^{\mathrm{th}}$ and 84$^{\mathrm{th}}$ quantiles. The median $\pm$ 1$\sigma$ values are reported in the title of each histogram. The latitude and longitude scales are in radians.}
	\label{fig:examplecorner}
\end{figure*}
To establish the priors for the parameters $v \sin{i}$, $\gamma$, and $i$, which should be consistent across all observations, we conducted a series of tests in which we allowed these parameters to vary within a range around their literature values. In particular, we kept $v \sin{i}$ fixed at 24.74 km s$^{-1}$, the value most frequently obtained in previous tests, and the inclination of the rotation axis to the line-of-sight $i$ at 90 degrees (a reasonable estimate derived from the value of v $\sin{i}$, R$_{\star}$ and P$_{\rm rot}$). Additionally, to break the degeneracy on the hemispheric location of the spots that arises when the inclination of the spin axis is 90 degrees, we assume that the spots are always located in the Northern hemisphere of the star. This star is observed close to the equator-on, such that a spot in the upper or, symmetrically, in the lower hemisphere of the star, with the same longitude and filling factor, contributes equally to the CCF profile. The spot radius, normalised to the stellar radius, was allowed to vary in the range [$10^{-2}$, 0.5] to prevent spots that were too small to be resolved by the mapping resolution, or too large and covering the entire stellar surface.  

In the second step, we used a model that accounted for the presence of two different \text{black} spots on the stellar surface  (hereafter, "Two-spots model") to fit the CCF. Note that this multi-spot model also considered cases where spots were entirely or partially overlapping. We adopted the same priors as the "One-spot model", which are detailed in Table \ref{tab:priorsSPOTS}. The logarithmic Bayesian evidence, $\log \mathcal{Z}$,  was used as an estimate of the goodness of the model. There is very strong evidence ($\Delta$ log $\mathcal{Z} \approx 2000 >> 5$) according to Kass-Raftery scale \citep{Kass10.2307/2291091}, supporting the "Two-spots model" for all the analysed CCF profiles of V1298 Tau.

Figure \ref{fig:exampleCCFmodel} shows an example of a CCF profile of V1298 Tau fitted with a "One-spot model" and "Two-spots model", and the corresponding spot configuration.  
As an example, Figure \ref{fig:examplecorner} shows the corner plot of the best-fit parameters acquired from fitting a CCF profile of V1298 Tau using the Two-spots model. Here, $\sigma_{\rm J}$ is the jitter noise, which refers to the noise level to be added to the model as an additional parameter to account for unmodelled contributions or other factors not considered by the model.

\section{Radial velocity time series analysis}\label{sec:analysis}
\begin{figure}[htp!]
  	\centering
  	\includegraphics[width=0.95\hsize]{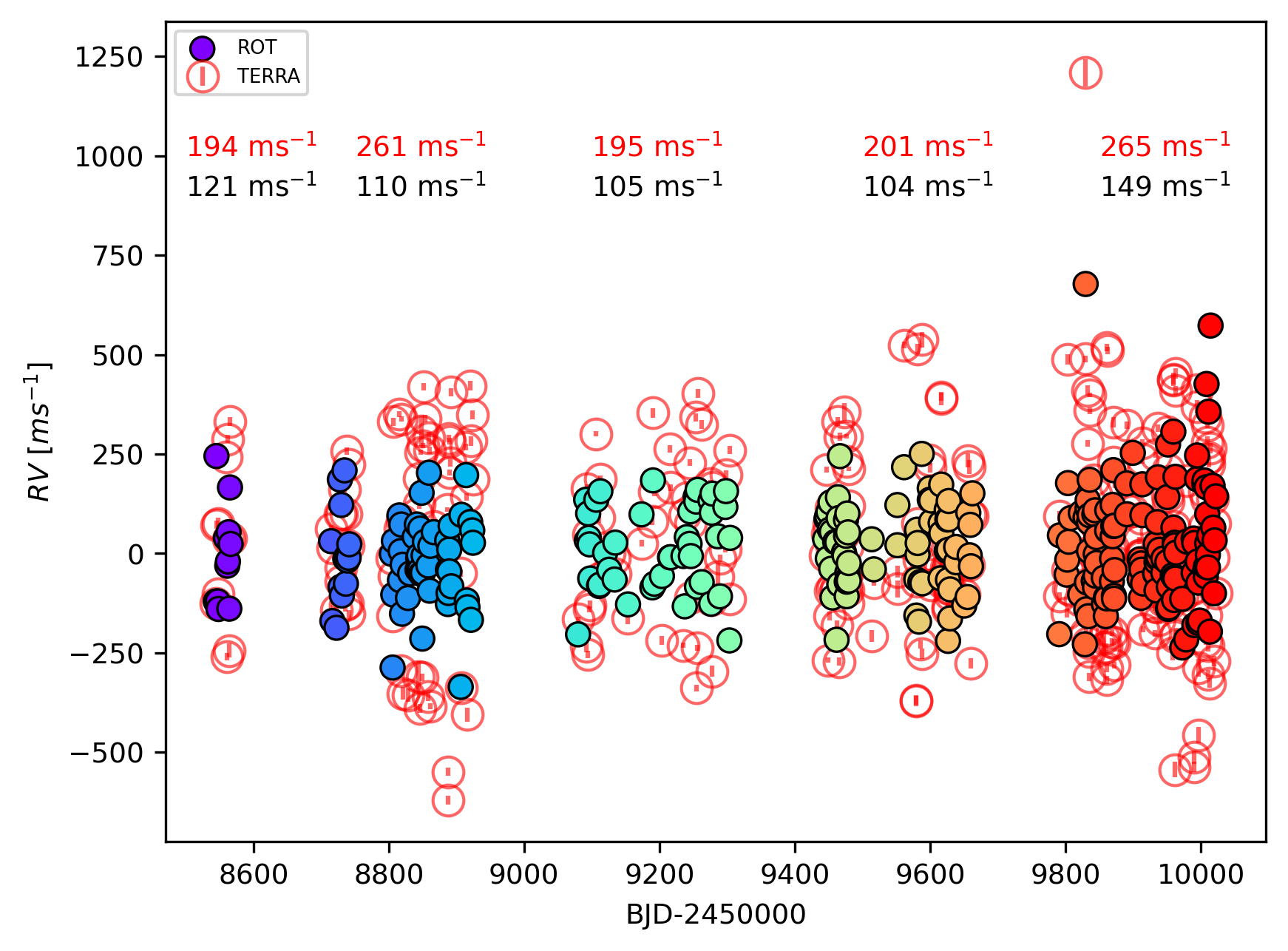}
  	\caption{Comparison between radial velocity measurements obtained with the \texttt{SpotCCF} (coloured filled points, ranging from blue to red as a function of the observation time) and TERRA RVs (red empty points). For each season we indicate the RMS of the corresponding RV subsample, in red the RMS of the RVs extracted from TERRA and in black the RMS of the \texttt{SpotCCF} RVs.}
  	\label{fig:SpotCCFvsTERRA}%
  \end{figure}

To extract the RVs of V1298 Tau we applied  \texttt{SpotCCF} to its CCF profiles, selecting the best model based on Bayesian evidence values, typically the Two-spots model.
We compared the RV time series obtained with \texttt{SpotCCF} to the RVs obtained with the TERRA pipeline (see Figure \ref{fig:SpotCCFvsTERRA}), since the latter shows a reduced RV dispersion compared to the DRS. 
The root mean square (RMS) of the RVs obtained with \texttt{SpotCCF} is significantly smaller (between 40\% and 60\%) than the RMS obtained with TERRA, with the largest decrease observed during the second season.  
In addition, we assessed the effect of including an extra spot in the \texttt{SpotCCF} model (Three-spots model) on the extraction of RVs. We verified that the Three-spots model exhibited greater RV dispersion compared to the Two-spots model in all observing seasons. This can be attributed to correlations between the RV measurements and the other parameters, such as $\gamma$ or the longitudes of the third spot, correlations that were not present in the Two-spots model (see an example in Figure \ref{fig:examplecorner2spot} and \ref{fig:examplecorner3spot}).

\begin{figure}[h!]
 \centering
            {\subfloat[\label{fig:glsSPOT}]{\includegraphics[width=0.95\hsize]{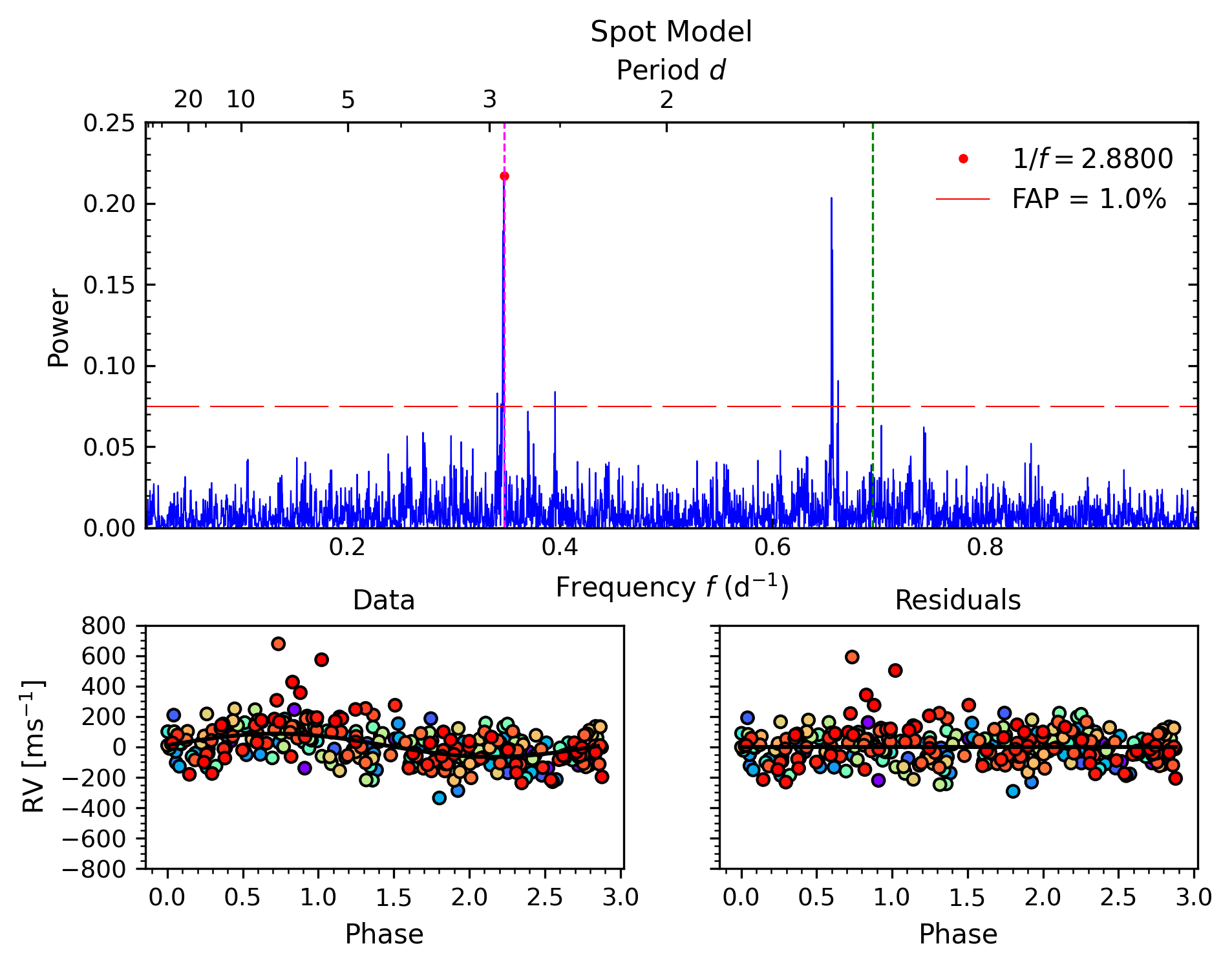}}\\
        \subfloat[\label{fig:glsTERRA}]{\includegraphics[width=0.95\hsize]{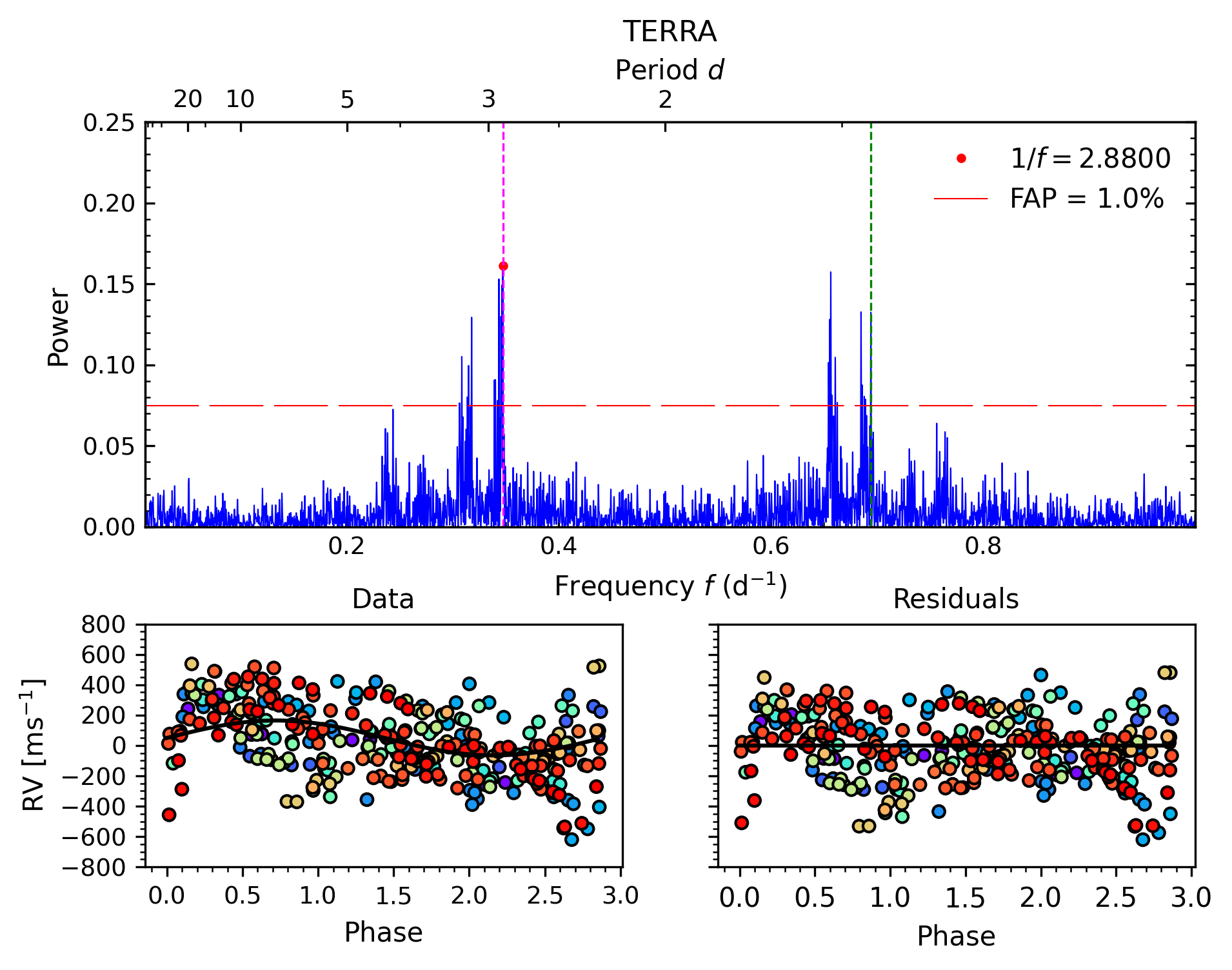}}
    }
    \caption{GLS periodogram of the V1298 Tau RVs obtained with the \texttt{SpotCCF} (panel a) and the TERRA pipeline (panel b). The red dots highlight the maximum power and the dashed red line indicates the FAP at 1.0\% (upper panels). The vertical dashed magenta line highlights the P$_{\rm rot}$ of the star, while the green dashed line indicates P$_{\rm rot}/2$. In the bottom panels, we showed the RVs (left) and the residuals (right) phase folded with the dominant period. The colour gradient, as shown in Figure \ref{fig:SpotCCFvsTERRA}, helps in identifying the modulation in the phase-fold residuals across different observing seasons.}
	\label{fig:glsSPOTeTERRA}
 \end{figure}

To search for periodicities in the RV data obtained with \texttt{SpotCCF}, we used the generalized Lomb-Scargle periodogram (GLS, \citealt{Zechmeister2009A&A...496..577Z}). 
The periodogram, see Figure \ref{fig:glsSPOT}, identifies a significant frequency at 0.34716 $\pm$ 0.00005 d$^{-1}$ (period of 2.8806 $\pm$ 0.0004 d), corresponding to the rotational period of the star, P$_{\rm rot}$, \text{highlighted with a dashed magenta line}, and the $1-f$ alias (frequency of about 0.65 d$^{-1}$). Figure \ref{fig:glsTERRA} shows the periodogram of the RVs obtained with the TERRA pipeline. In this case, as well, the periodogram identifies a significant frequency at the P$_{\rm rot}$ of the star. However, the frequency at the first harmonic of the P$_{\rm rot}$ of the star (green dashed line), which is present in the TERRA periodogram, disappears in the \texttt{SpotCCF} periodogram. Furthermore, it is worth noting that the \texttt{SpotCCF} periodogram does not reveal the intricate pattern of peaks around the P$_{\rm rot}$ and its harmonics, respectively, which are clearly visible in the TERRA periodogram. The results of the sinusoidal fit from both periodograms are summarised in Table \ref{table:GLS_SPOTeTERRA}.
 \begin{figure*}[h!]
    \resizebox{\hsize}{!}
	{\subfloat[\label{fig:prewSPOT}]       {\includegraphics[width=0.5\hsize]{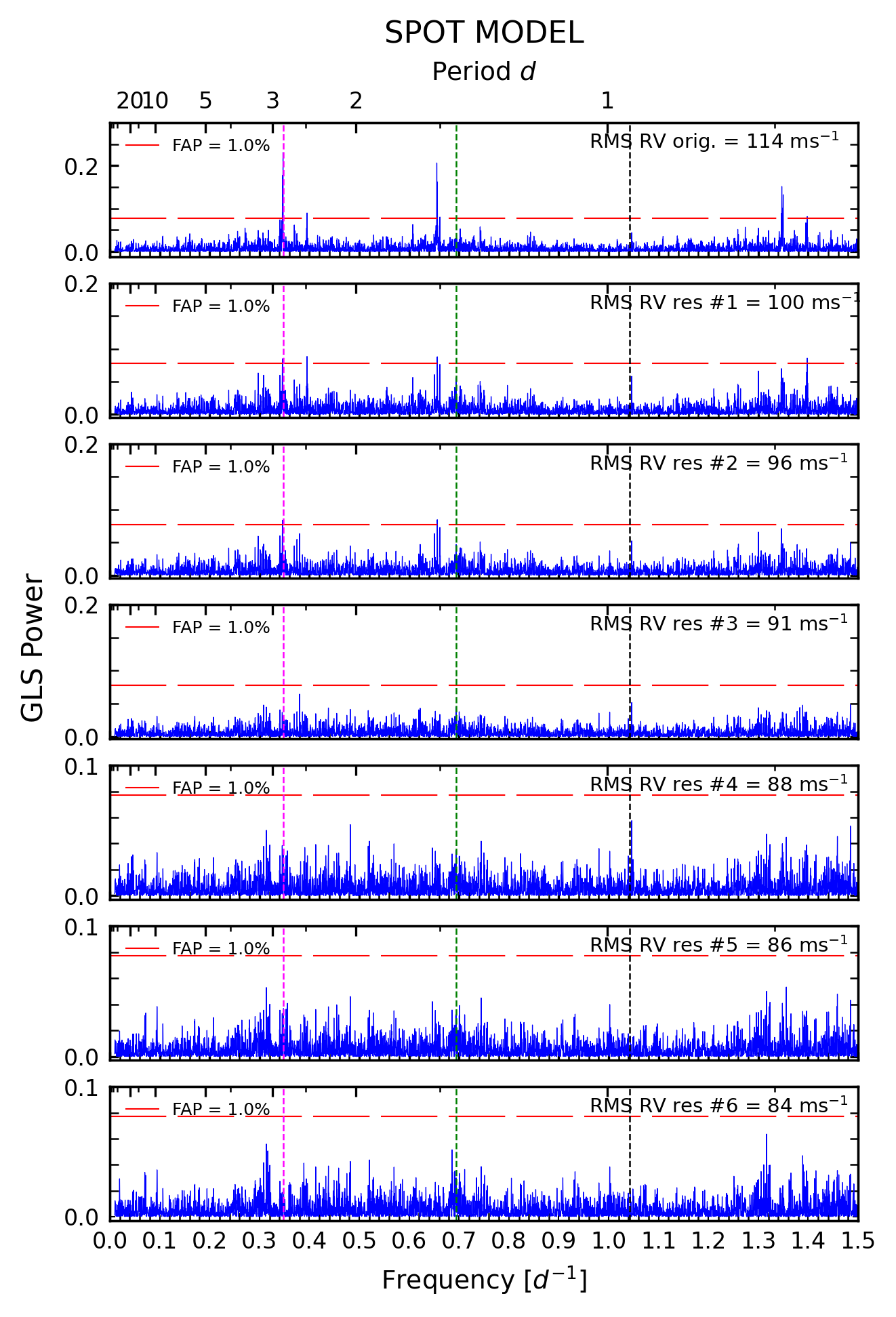}} 
		\subfloat[\label{fig:prewTERRA}]{\includegraphics[width=0.5\hsize]{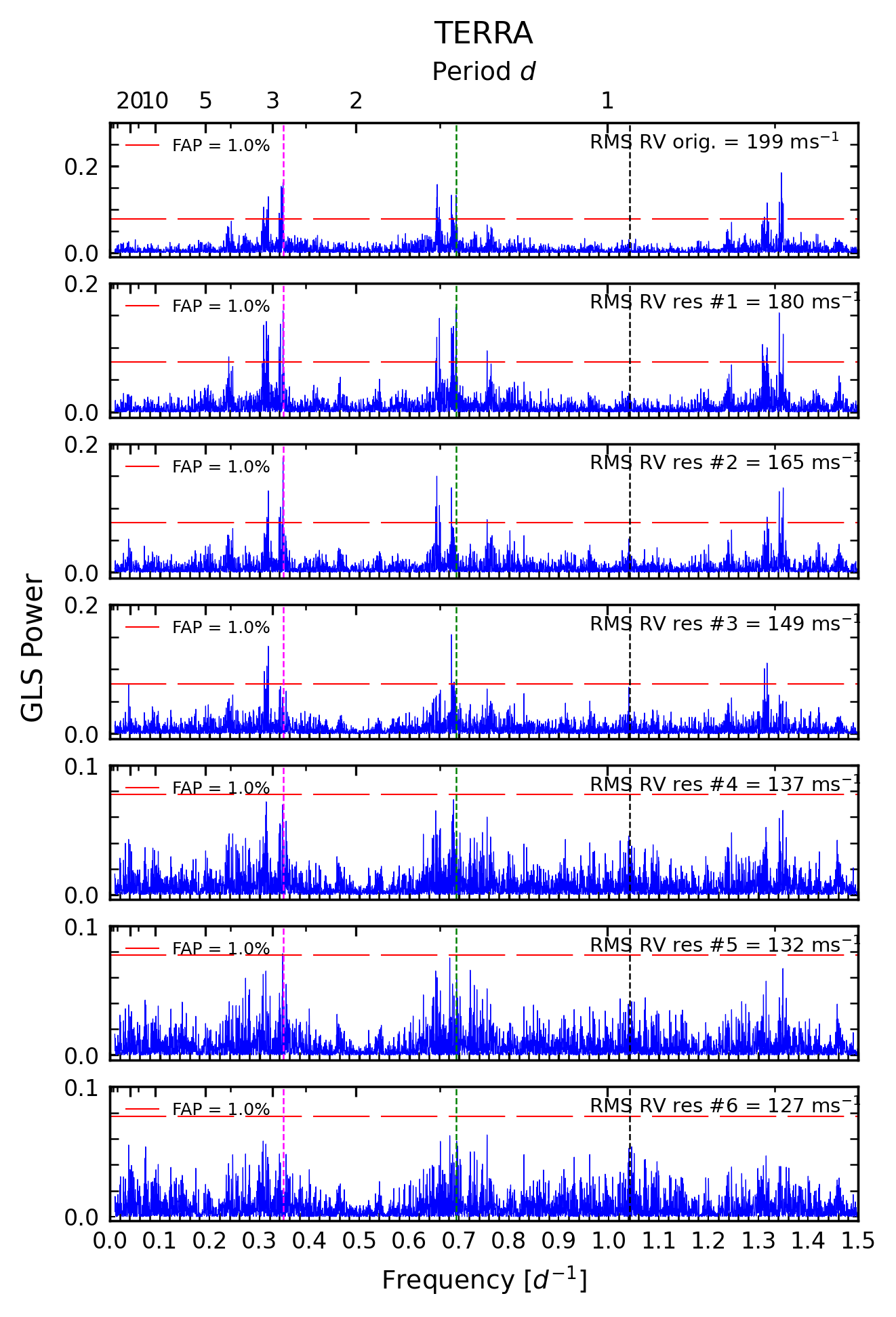}} 		
	}
	\caption{GLS periodogram of the V1298 Tau RVs obtained with the \texttt{SpotCCF} (left panels) and the TERRA pipeline (right panels) and their residuals after recursive pre-whitening. For each panel: the vertical and dashed lines indicate the stellar rotation frequency (in magenta) and its harmonics (in green and black); the horizontal red dashed line highlights the FAP at 1.0\%. Each panel reports the RMS of the dataset.}
	\label{fig:prewhitening}
\end{figure*}

\begin{table}[h!] \scriptsize
	\caption{Summary of the sinusoidal fits obtained from the periodograms applied to the V1298 Tau RVs derived with \texttt{SpotCCF} and TERRA.}   
	\label{table:GLS_SPOTeTERRA} 
	\centering             
	\begin{tabular}{lcc}   
		\toprule[0.05cm]
		\toprule
		Parameter & \texttt{SpotCCF} & TERRA  \\
		\midrule            
		\enskip
		Weighted RMS of dataset (m s$^{-1}$)& 115.29 & 199.27 \\ 
		\enskip
		RMS of residuals (m s$^{-1}$) & 101.12 & 182.49 \\
		\enskip
		Mean weighted internal error (m s$^{-1}$) & 6.08 & 8.79\\
		\enskip 
		Best sine frequency (d$^{-1}$)& 0.34716 $\pm$ 0.00005 & 0.34722 $\pm$ 0.00006 \\
		\enskip 
		Best sine period (d) & 2.8806 $\pm$ 0.0004 & 2.8800 $\pm$ 0.0005 \\
		\enskip 
		Amplitude (m s$^{-1}$) & 79 $\pm$ 8 & 115 $\pm$ 15 \\
		\bottomrule[0.05cm]                
	\end{tabular}
\end{table}

We noted that the RV semi-amplitude obtained from \texttt{SpotCCF} is about 30\% lower than the one obtained with TERRA, and the RMS of the residuals obtained from the subtraction of the best sinusoid from \texttt{SpotCCF} RVs is about  45\% lower than the TERRA RMS of the residuals. 
Furthermore, it can be noted that the modulation present in the phase-folded residuals of TERRA disappears when \texttt{SpotCCF} is used (see the bottom right panel in Figures \ref{fig:glsSPOT} and \ref{fig:glsTERRA}). This is also confirmed by the periodogram, as the \texttt{SpotCCF} periodogram appears cleaner and the rotational period seems better identified.
This result suggests a substantial removal of rotation-related frequencies by \texttt{SpotCCF}. The presence of a peak at the P$_{\rm rot}$ that remains visible could be attributed to the approximate representation of spots, both in terms of their number and shape, within the \texttt{SpotCCF} model, or to other phenomena responsible of the stellar variations.

We performed an additional test by calculating the GLS periodograms for both the original data and residuals obtained after recursive pre-whitening for both datasets. The periodogram of TERRA RVs exhibits a complex structure centred around the stellar rotation frequency, with signals related to the stellar activity (at the rotational frequency or its harmonics) prominently dominating the periodograms even after four iterations of pre-whitening. However, the periodogram of the \texttt{SpotCCF} RVs does not display harmonics of the rotation frequency, and the signal at the stellar rotational frequency disappears after the second pre-whitening step (see Figure \ref{fig:prewhitening}).

Our analysis strongly supports that the \texttt{SpotCCF} model efficiently removes part of the contribution from the stellar activity related to the rotation.

\section{Spot characterisation}\label{sec:spotanalysis}
The fit of the CCFs profile with \texttt{SpotCCF}  allows for characterising the spots present on the stellar surface. In particular, latitude and longitude provide the position of the spot, while the projected filling factor (ff$_{\rm p}$) suggests its size (see Table \ref{tab:params}).  

From a preliminary analysis, we decided to make the raw assumption that the spot system consists of a bigger (Spot A) and a smaller (Spot B) spot evolving separately. 

   \begin{figure}[htp!]
  	\centering
  	\includegraphics[width=0.99\hsize]{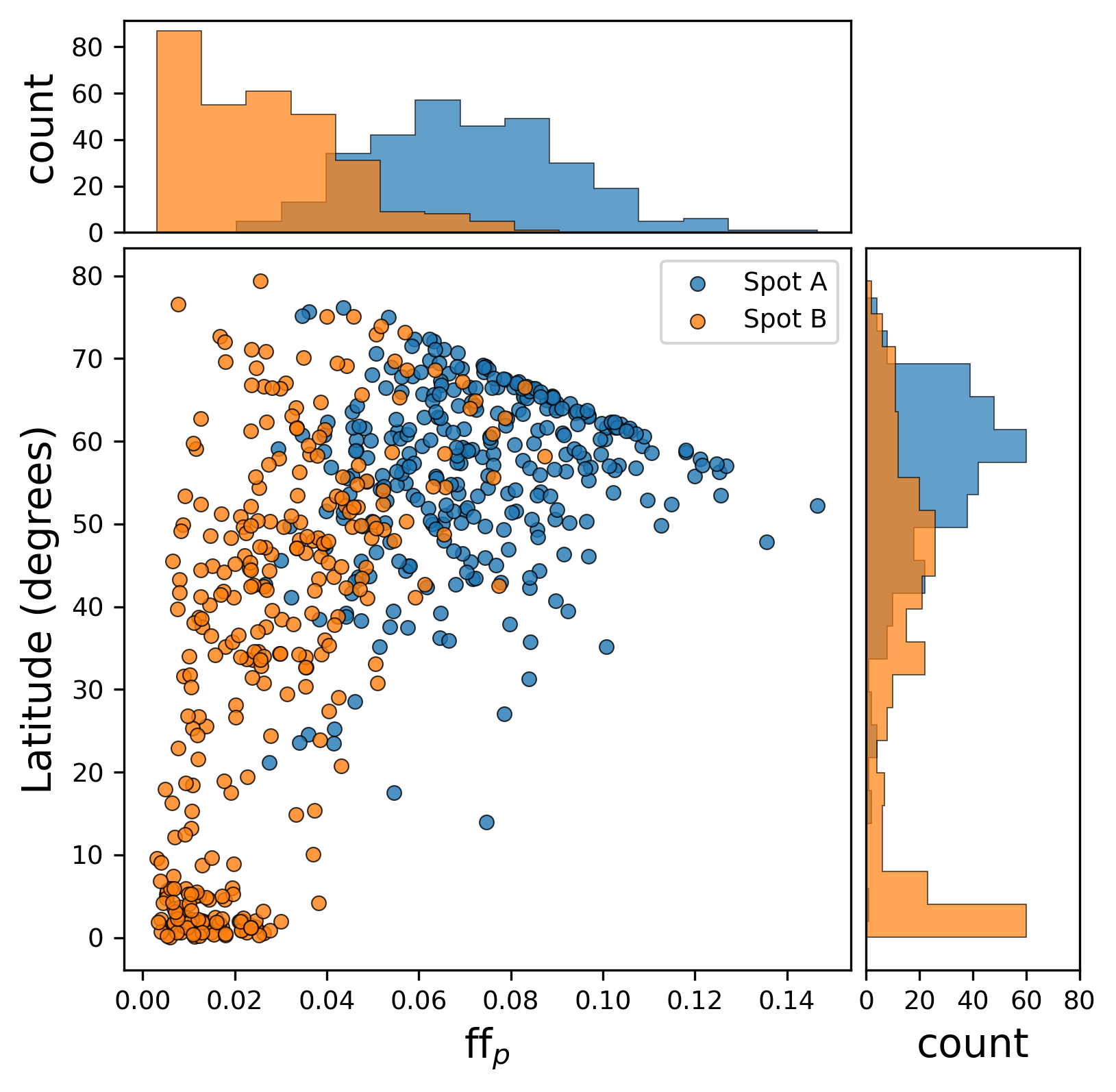}
  	\caption{The main plot shows the relationship between latitude and ff$_p$ for spots on V1298 Tau. Blue points represent the spot with the highest ff$_{\rm p}$ (Spot A), while orange points correspond to spots with the lowest ff$_{\rm p}$ (Spot B). The histograms on the top and right side of the main plot depict the distributions of latitudes and ff$_p$, respectively.}
  	\label{fig:latvsffp}
  \end{figure}
In Figure \ref{fig:latvsffp} we show the relation between the latitude of the spots and the projected filling factor obtained for V1298 Tau, and the respective distributions (see the histograms on the top and right side of the main plot). 
According to our assumption, we found an indication of two main peaks, with larger spots (ff$_{\rm p}$ > 0.06) preferentially in the range ($\approx$ 45-90) degrees, and smaller spots (ff$_{\rm p}$ < 0.02) in the range (0-10) degrees.  The linear upper envelope that characterises the larger spots is attributable to the upper limit of the spot radius boundary. 

 \begin{figure}[htp!]
	\centering
	\includegraphics[width=\hsize]{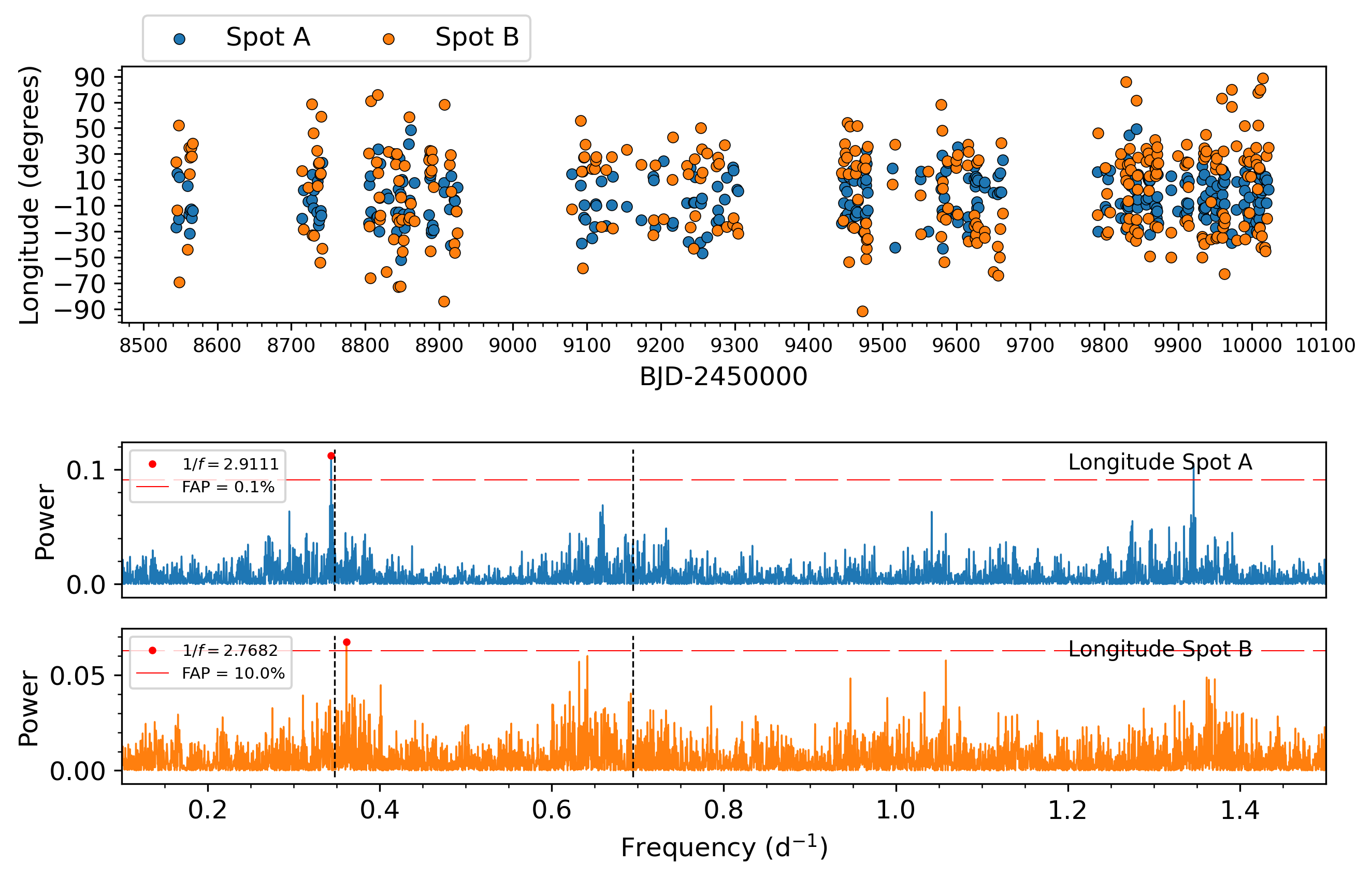}
	\caption{(Upper panel) V1298 Tau longitude time-series. The blue points indicate the large spots (Spot A), while the orange points are the small spots (Spot B). (Lower panels) GLS periodogram performed for the longitude values obtained for Spot A and Spot B, respectively. The red horizontal line indicates the false alarm probability (FAP) level at 0.1\% and 10\%, respectively. The black dashed lines highlight the rotational period of the star and its harmonic.}
	\label{fig:longitudeperiod}%
\end{figure}

We performed the GLS periodogram analysis of longitudes and projected filling factor obtained for V1298 Tau.

In the upper panel of Figure \ref{fig:longitudeperiod} we show the longitude time series for Spot A and Spot B, in blue and orange respectively, and the lower panels show the GLS periodograms performed for each spot, respectively. 
A significant peak at 2.91 days, near the rotational period of the star is identified for Spot A with a false alarm probability (FAP) $\leq$ 0.1\%, while a significant peak (FAP $\leq$ 10\%) at 2.77 days is identified for  Spot B. 

 \begin{figure}[h!]
	\centering
	\includegraphics[width=\hsize]{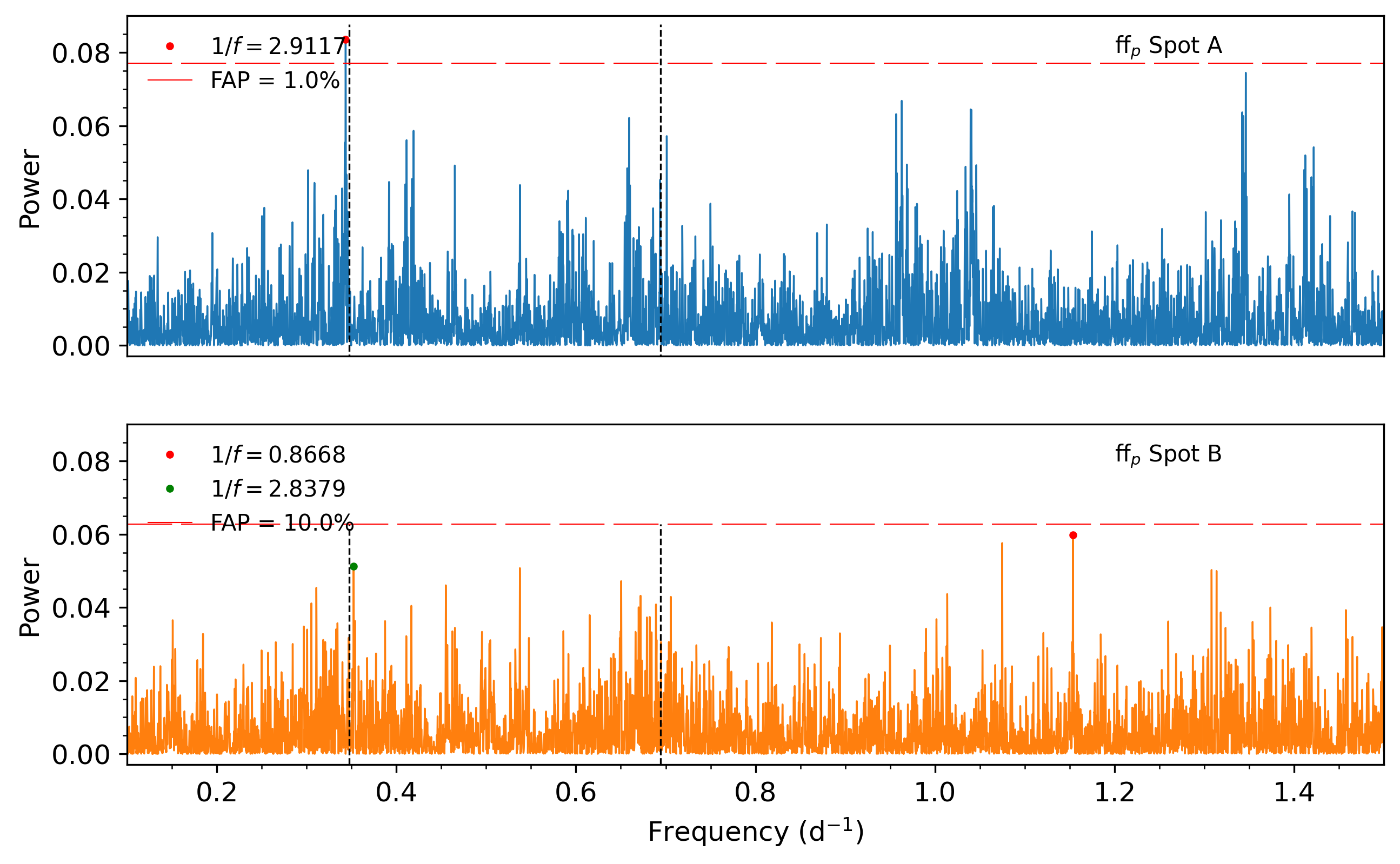}
	\caption{GLS periodogram of the projected filling factor of Spot A (upper panel) and Spot B (lower panel). The red horizontal line indicates the false alarm probability (FAP) level at 1\% and 10\%, respectively. The black dashed lines highlight the rotational period of the star and its harmonic.}
	\label{fig:ffpGLS}%
\end{figure}

The GLS periodogram of the projected filling factor of the two spot distributions is shown in Figure  \ref{fig:ffpGLS}. Even in this case, a significant peak (FAP $\leq$ 1\%) is identified for Spot A at 2.91 days, while Spot B shows a lower significant peak (FAP $\approx$ 10\%, 2.84 days).  

Both periodogram analyses indicate that the deformations in the CCF profiles are modulated with the stellar rotation, confirming our hypothesis that these deformations are caused by the presence of stellar spots rotating solidly with the star.

The lower significance of the peaks observed for Spot B is not surprising, as Spot A is larger and therefore contributes more significantly to the CCF profile, making it easier to detect. However, since Spot B is smaller, when there are more than two spots on the stellar surface or when a simplistic description of a circular spot is not sufficient, there can be multiple configurations for Spot B. Furthermore, the definition of "large" and "small" spots are not strictly defined and are relative to the specific configuration, as there can be cases where both spots should be categorized as "large", but one is slightly smaller than the other, though still falling within the "small" category. 

\begin{figure}[h!]
	\centering
        \includegraphics[width=0.98\hsize]{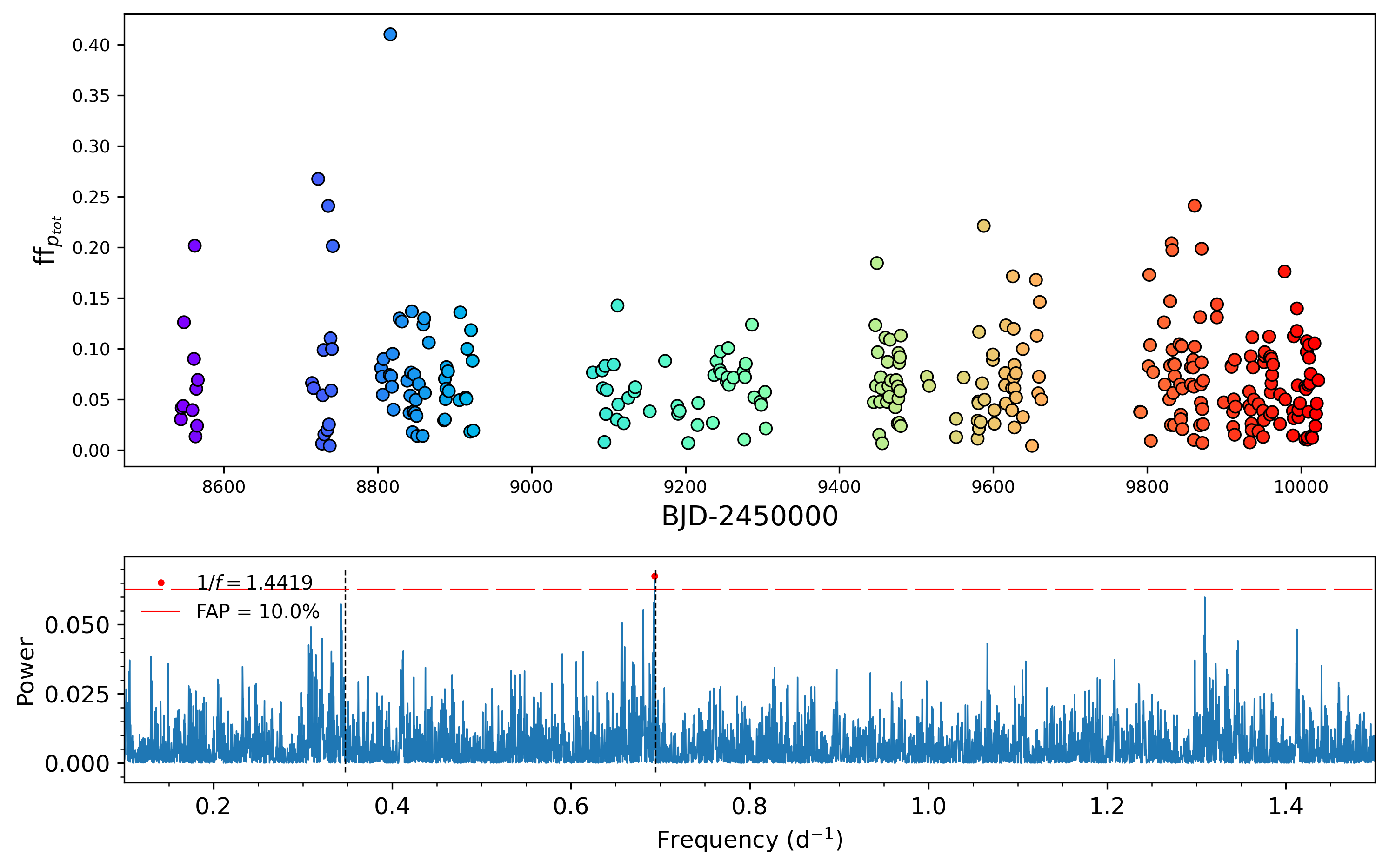}
     \caption{Total projected filling factor time-series of V1298 Tau (upper panel) and corresponding GLS periodogram (lower panel). The points are coloured following a colour scale from blue to red as a function of the observation time. The red horizontal line in the periodogram indicates the FAP level at  10\%, while the black dashed lines highlight the rotational period of the star and its harmonic.}
	\label{fig:ffptot}
\end{figure}

Furthermore, we analysed the total area covered by the spots, ff$_{\rm p_{\rm tot}}$, which was obtained by summing up the ff$_{\rm p}$ of each individual spot. The upper panel of Figure \ref{fig:ffptot} shows the ff$_{\rm p_{\rm tot}}$ times-series. 

It is worth noting that the range of the total projected filling factor obtained for V1298 Tau is consistent with the value proposed by \citet{Messina2001A&A...366..215M}, according to which, a star like V1298 Tau, with a rotational period of about 3 days, has approximately 15\% of its surface covered by spots.  

The GLS periodogram of the ff$_{\rm p_{\rm tot}}$ time-series reveals a peak corresponding to the second harmonic of the P$_{\rm rot}$ of the star (FAP $\leq 10$\%). 
Since we are monitoring the total area covered by the spot, the presence of peaks at both P$_{\rm rot}$ and P$_{\rm rot}/2$ indicates a non-uniform spot distribution, concentrated along two distinct mean longitudes separated by 180 degrees. Moreover, the different power of these peaks suggests that these two opposing spot distributions cover different areas on the stellar surface. 
This behaviour appears to remain fairly stable over time.

As a further validation of the method, we compare the spot configuration of observations obtained on the same nights. 
In Figure \ref{fig:spots} we present the spot configurations obtained from pairs of observations taken at very few hours of separation. Specifically, we compare the spot configurations obtained from observations taken during the same night, but a few hours apart. It is important to note that each observation was analysed independently, and obtaining the same configuration of spots on the same night (with very few exceptions in 2023/01/16 and 2023/01/17) serves as a confirmation of the reliability of the method.

\section{Differential stellar rotation}\label{sec:diffrot}
To test the presence of differential rotation, we categorised the spots into two groups: larger spots located between latitudes 60-90 degrees (referred as Spot A at $\geq$ 60$^\circ$), and smaller spots located between latitudes 0-40 degrees (Spot B at $\leq$ 40$^\circ$).  This eliminates the "small" spots at high latitudes. 

\begin{figure}[h!]
	\centering
	\includegraphics[width=\hsize]{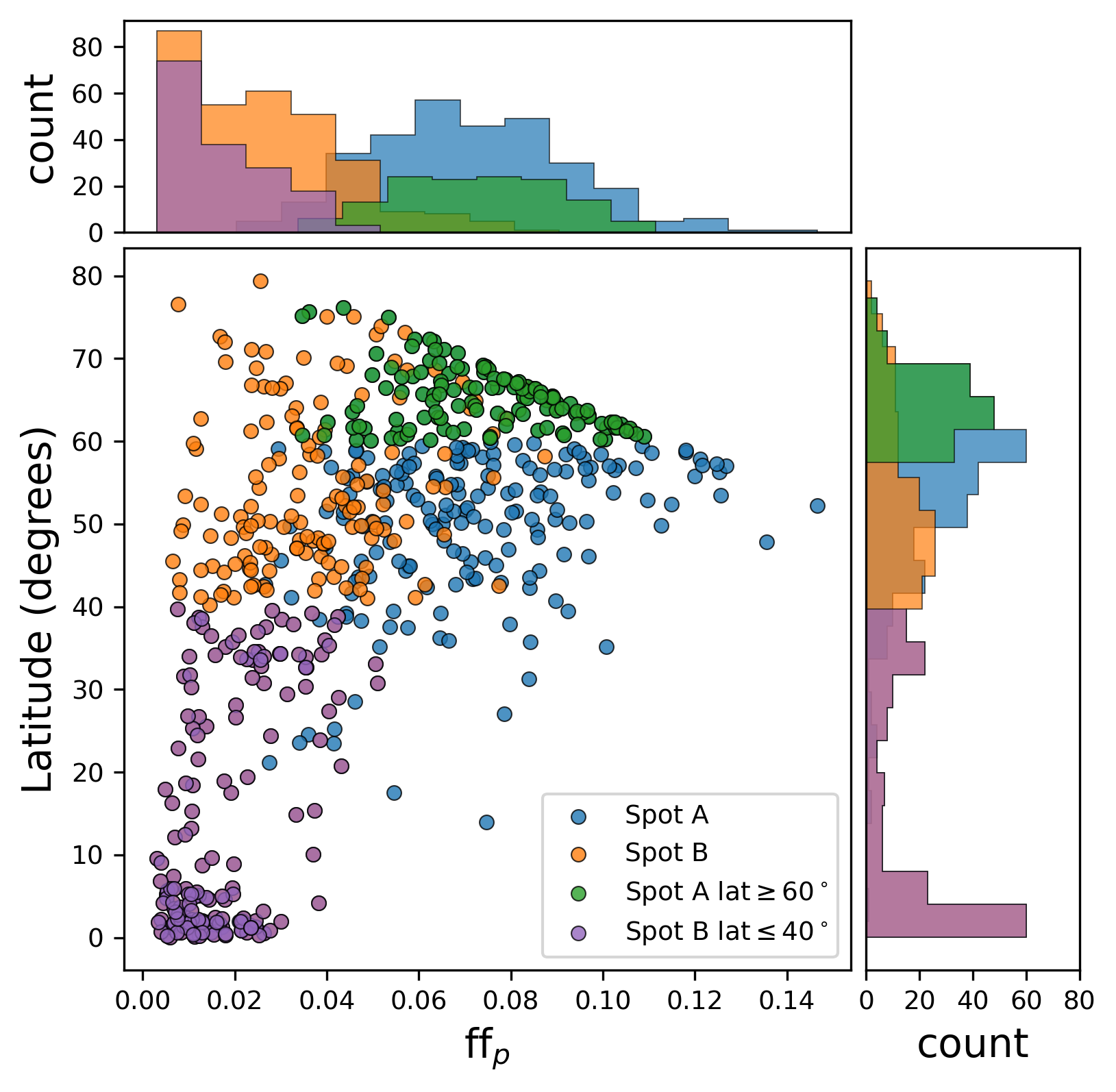}
	\caption{The main plot shows the relationship between latitude and ff$_{\rm p}$. The histograms on the top and right side of the main plot depict the distributions of latitudes and ff$_{\rm p}$, respectively. Blue points represent the spot with the highest ff$_{\rm p}$ (Spot A), while orange points indicate the spots with the lowest ff$_{\rm p}$ (Spot B). Purple points and distribution represent Spot B with latitude values lower than 40 degrees, while green points and distribution indicate Spot A with latitude values higher than 60 degrees.}
	\label{fig:latselection}%
\end{figure}

Figure \ref{fig:latselection} illustrates this new selection, depicting the distribution of latitude as a function of the ff$_{\rm p}$. 

\begin{figure}[h!]
	\centering
	\includegraphics[width=\hsize]{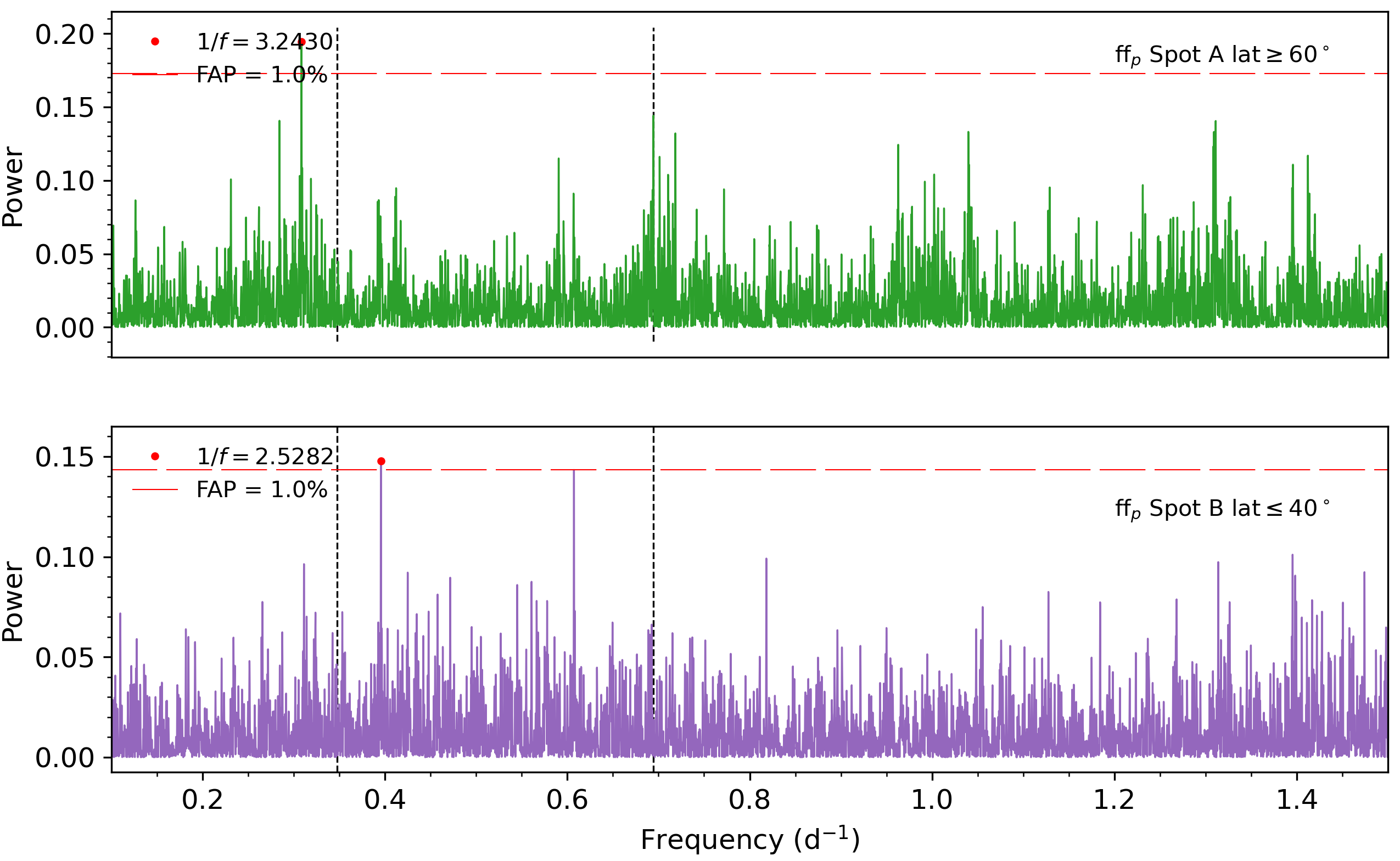}
	\caption{GLS periodogram of the projected filling factor of Spot A with a latitude higher than 60 degrees (upper panel) and Spot B with a latitude lower than 40 degrees (lower panel). The red horizontal line indicates the false alarm probability (FAP) level at 1\%. The black dashed lines highlight the rotational period of the star and its harmonic.}
	\label{fig:ffp_latselection}%
\end{figure}

The GLS periodogram was performed on the projected filling factor distribution of the two spots (see Figure \ref{fig:ffp_latselection}). The analysis revealed a significant peak (FAP $\leq$ 1\%) for Spot A and Spot B, at periods of 3.24 and 2.53 days, respectively. These periods may provide a suggestion of a differential rotational velocity of the star, with a higher velocity at lower latitudes (P = 2.53 days close to the equator) and lower rotational velocity at higher latitudes (P = 3.24 days close to the pole). Additionally, it can be observed that the peak at 3.24 days is one of the peaks near the rotational period of the star in the TERRA periodogram, which was removed by the \texttt{SpotCCF} method. 

\section{Detection sensitivity by direct injection of the planetary signal into the data} \label{sec:injection}
 In order to obtain further validation of the \texttt{SpotCCF} method we tested the ability to recover the planetary signal in the time series of V1298 Tau. For this purpose, we created different datasets by injecting a planetary signal with an orbital period of 4.9 days\footnote{Please note that the injected planet has a different period than the transiting planets, which
enables good control over the amplitude of the signal.} and using different amplitudes (K = 37, 75, 100, 150 m s$^{-1}$) in the RV dataset obtained with \texttt{SpotCCF} and with the TERRA pipeline, respectively. 
 Before proceeding, we tested the reliability of the method by injecting the signal on the CCFs verifying that the two procedures give the same radial velocity. To simulate a planetary signal, the CCFs were blue or redshifted with the desired amplitude, period and phase. We performed the fit for all the shifted CCFs and the results were compatible with the fit obtained with the original CCFs. For this reason, and since
the fit of CCFs is computationally demanding, we chose to inject the planetary signal directly on the original RVs time series.

 \begin{figure*}[htp!]
 	\centering
 	\includegraphics[width=0.97\textwidth]{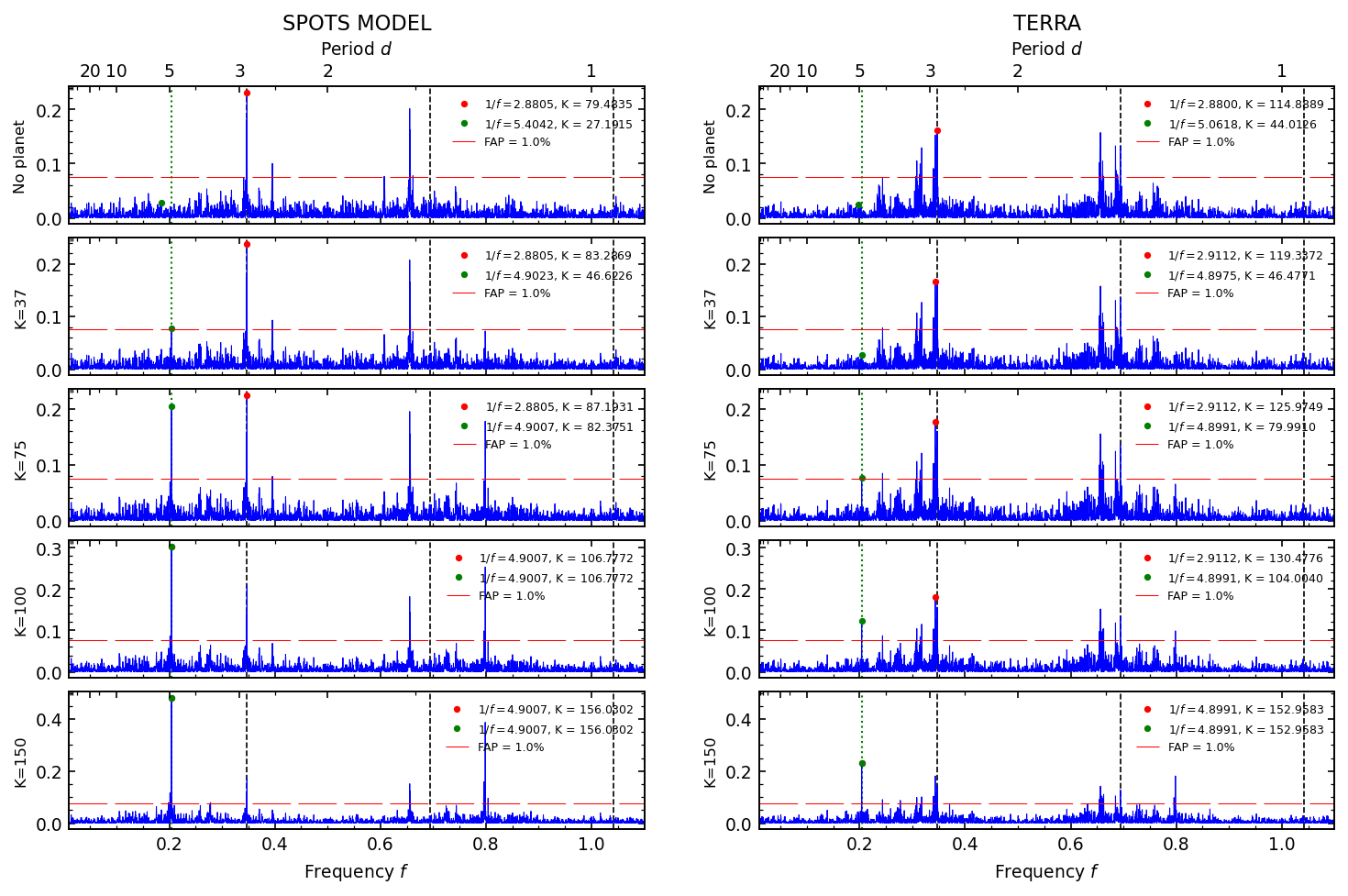}
 	\caption[GLS periodogram of the V1298 Tau RVs obtained with the \texttt{SpotCCF} model and the TERRA pipeline where we injected a planetary signal to test the detection sensitivity]{Comparison between the GLS periodograms of the V1298 Tau RVs obtained with the \texttt{SpotCCF} (left panels) and the TERRA pipeline (right panels) where we injected a planetary signal at P$_{\rm orb}$ = 4.9 d, with different amplitude (from top to bottom K=0, 37, 75, 100, 150 m s$^{-1}$), the vertical and dashed black lines indicate the stellar rotation frequency and its harmonics; the green line marks the orbital frequency of the injected planet, while the red dashed horizontal line highlights the FAP at 1.0\%. The upper panel shows the original dataset without planet injection.}
 	\label{fig:injplanet}%
 \end{figure*}
 
 The GLS periodogram was computed for each dataset. Figure \ref{fig:injplanet} shows the comparison between the periodograms of the  RVs+planet (injected at different amplitude) obtained from the original \texttt{SpotCCF} RVs (left panel) and from the original TERRA  dataset (right panel). In both datasets (\texttt{SpotCCF} and TERRA) the
 GLS periodogram finds a significant peak at the rotational period of the star. From the second row of the figure, it can be seen the periodogram for the RVs with the planet injection at different amplitudes. 

 The GLS periodogram finds a significant peak (FAP $<$ 1.0\%) at the orbital period of the injected planet with K $\approx$ 37 m s$^{-1}$ (corresponding to $M \sin{i} \approx 0.35$ $M_{\rm J}$ ) for the RVs obtained with \texttt{SpotCCF}. An analogous peak is significant for the TERRA RVs only for an injected planet with K $\approx$ 75 m s$^{-1}$ (corresponding to M $\sin{i}$ $\approx$ 0.70 M$_{\rm J}$ ). 

  This test suggests that using this method, it is possible to reliably detect a lower-massive planet, with an amplitude of the injected signal lower than what is required for the signal to be identified in the TERRA dataset. This result is consistent with the finding that the \texttt{SpotCCF} model reduces systematic effects caused by stellar activity.

 \section{Discussion and conclusions}\label{sec:summary}
In this work, we described \texttt{SpotCCF}, which is a stellar photosphere model fit. It is able to extract the spot configuration of the star and also optimise the radial velocities extraction in young-active stars, based on the cross-correlation function technique. This model takes into account the deformations of the CCF due to the presence of multiple spots on the stellar disc, in the presence of significant rotation. 

  To test the validity of our model, we analysed V1298 Tau HARPS-N observations, which exhibit distorted CCF profiles due to the high-activity level of this target. 
 The original CCF profiles produced by the DRS pipeline were corrected for anomalous deformations present in the wings and in the core of the line, as well as in the CCF$_{\rm B}$ of the sky spectrum obtained using fibre B. 

Then, we analysed over 300 HARPS-N observations of V1298 Tau using a Two-spots model.
From the parameters through the fit performed on the CCF profiles, we extracted information about the spots, including latitude, longitude and the area covered by each spot. 
We found two main distributions with polar/high-latitude spots, in agreement with the high rotation of the star, and smaller low-latitude spots \citep[e.g]{Barnes2001MNRAS.326.1057B, Strassmeier2002AN....323..309S, Cang2021A&A...654A..42C}. 
To analyse the spot properties, we separated the high-latitude spots (Spot A, ff$_{\rm p}$ $\geq$ 0.04 and latitude $\geq$ 60$^\circ$) and the smaller low-latitude spots (Spot B, ff$_{\rm p}$ < 0.04 and latitude $\leq$ 50$^\circ$). Under this assumption, the GLS periodogram of the filling factor of Spot A and Spot B showed significant peaks (FAP $\leq$ 1\%) at 3.24 and 2.53 days, respectively, suggesting a solar-like differential rotation of the star, $\alpha = (\Omega_{\rm eq} - \Omega_{\rm pol})/\Omega_{\rm eq} \simeq 0.2$ ($\alpha_{\odot} = 0.2$ \citealt{Balona2016MNRAS.461..497B}), with lower rotation at higher latitudes. This estimate places this target in the uppermost region of Figure 11 provided by \citet{Reinhold2013A&A...560A...4R}, confirming the extreme behaviour of this star.
  
The average total area covered by the spots is consistent with the range proposed by \citet{Messina2001A&A...366..215M}.  The GLS periodogram of the total projected filling factor, ff$_{\rm p_{\rm tot}}$, showed a significant peak at the P$_{\rm rot}$/2 with a FAP $\leq$ 10\%, indicating that the features identified in the CCF profiles are actually effects of few inhomogeneities (spots) on opposite hemispheres of the stellar surface modulated by the rotation. 

The consistency of the spot configuration obtained from different observations taken during the same night, but a few hours apart, confirms the reliability of the method.

Moreover, our results are consistent with the spot configuration of V1298 Tau obtained by \citet{Morris2020ApJ...893...67M}. In that paper, the author modelled the K2 Campaign 4 light-curves of V1298 Tau, identifying a three-spots configuration. Two main spots were located in the range 60-90 degrees (northern hemisphere, NH) and 30-60 degrees (southern hemisphere, SH), respectively, while a smaller spot was positioned at lower latitudes (0-30 NH). A starspot covered fraction, $f_s$, was estimated at 9$^{+1}_{-2}$ per cent. 
The observed spot configuration aligns with our two spot distributions, featuring a larger spot up to 60 degrees and a smaller one in the range 0-40 degrees.
The total projected filling factor of V1298 Tau estimated by using \texttt{SpotCCF} is consistent with their estimation. The median value of $ff_{p{\rm tot}}$ of 0.06$\pm$ 0.04 obtained by \texttt{SpotCCF} considers only the spot coverage of the visible hemisphere, representing half of the $f_s$ estimated by \citet{Morris2020ApJ...893...67M}.

The \texttt{SpotCCF} model applied to the HARPS-N observations of V1298 Tau enables the reduction of the contribution of stellar activity affecting the RVs measurements.
We compared these RVs with those obtained using the TERRA pipeline to highlight the benefits of the proposed method. We observed that the RVs obtained with \texttt{SpotCCF} model showed lower dispersion, with a decrease ranging from 40\% to 60\% in each season, compared to the TERRA dataset. Additionally, a search for periodicities in the RV dataset revealed a significant peak at the rotational period of the star, with reduced RV amplitude for the \texttt{SpotCCF} RVs (about 30\% lower than those obtained with TERRA) and reduced dispersion in the residuals. These results suggest that the new method for RV extraction partially mitigates the contribution of stellar activity modulated with stellar rotation. This conclusion is further supported by the GLS periodogram performed on the \texttt{SpotCCF} dataset and on its residuals obtained after recursive pre-whitening, which show the removal of the peak at the harmonics of the rotation after the first pre-whitening, while they are still present in the periodogram of the TERRA dataset. 
These results, particularly the significant peak at P$_{\rm rot}$ persisting even after SpotCCF modelling, imply that stellar variations are not fully explained by the existence of two prominent spots on the stellar surface alone.  The variability observed in the original RV time-series results not only from the presence of these two major spots but also from additional factors not accounted for in the model. These factors may include phenomena like the presence of plages or unmodelled spots, smaller than those significantly impacting the CCF profile. Additionally, there may be unmodelled effects associated with larger spots, such as the inhibition of convection, leading to changes in the convection blueshift when the spots are visible \citep{Cavallini1985A&A...143..116C, Dravins1981A&A....96..345D, Meunier2015A&A...583A.118M, Meunier2017}.
We also tested the detection sensitivity of the method by directly injecting a hypothetical planetary signal (P$_{\rm orb} = 4.9$ days) into the data. The results obtained from the GLS periodogram suggest that from the RV times series obtained with \texttt{SpotCCF} method, a planet with an amplitude of the injected signal lower (K $\approx$ 37 m s$^{-1}$) than that necessary for the signal to be identified in the TERRA dataset (K $\geq$ 75 m s$^{-1}$), can be reliably detected.

All the results of this work confirm that the developed method can extract information about the spot configuration of the star and also optimise the radial velocity extraction in young/active stars, improving sensitivity and ability to recover planetary signals and reducing the probability of identifying signals that are actually due to stellar activity but can be mistaken as planetary signals.

We plan to apply the method proposed in this work to other targets that exhibit high values of $v \sin{i}$, in order to assess the applicability range of the model. Furthermore, we could apply \texttt{SpotCCF} model to the single spectral lines that are predominantly affected by rotational broadening and distorted by the presence of spots, in order to more accurately characterise the line profiles.
    
\begin{acknowledgements}
      We thank the anonymous referee for her/his helpful comments and suggestions. 
      We acknowledge partial support of Ariel ASI-INAF agreement no. 2021-5-HH.0. 
      We acknowledge PRACE for awarding access to the Fenix Infrastructure resources at CINECA, which are partially funded from the European Union’s Horizon 2020 research and innovation programme through the ICEI project under the grant agreement No. 800858.
      The research activities described in this paper were carried out with contribution of the Next Generation EU funds within the National Recovery and Resilience Plan (PNRR), Mission 4 - Education and Research, Component 2 - From Research to Business (M4C2), Investment Line 3.1 - Strengthening and creation of Research Infrastructures, Project IR0000034 – “STILES - Strengthening the Italian Leadership in ELT and SKA”.
\end{acknowledgements}

    \balance{
    \bibliographystyle{aa}
    \bibliography{bibliography_YO01}}

   \begin{appendix}
   \section{Spot model}\label{appendice:modello}
\begin{figure}[h!]
	\centering
	\includegraphics[scale=0.17]{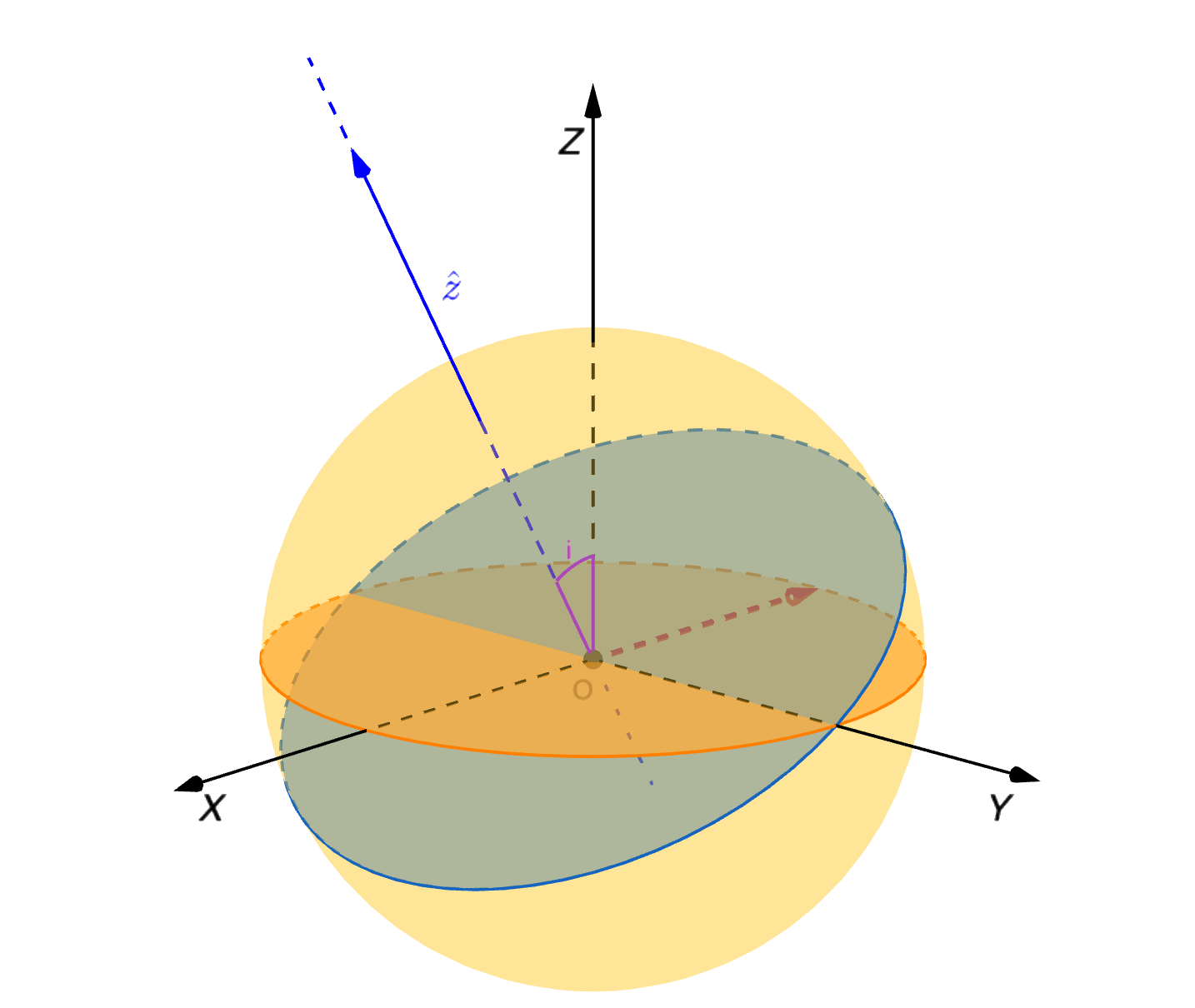}
	\caption{The $OXYZ$ reference frame with its origin in the barycentre $O$ of the star, the $Z$ axis along the stellar spin axis $\Omega_{\star}$, and the $XZ$ plane oriented as to contain the line of sight $\hat{z}$. The inclination of the stellar spin axis to the line of sight is indicated with $i$. The red vector indicates the projection of the rotation axis over the plane of the sky, that is the plane $xy$ of the reference frame $Oxyz$, obtained from $OXYZ$ by making a rotation of the angle $i$ along the $Y$ axis. Note that the projection of the spin axis onto the $xy$ plane is opposite to the positive direction of the $x$ axis because the angle $i$ is taken positive (counter-clockwise) from the spin axis to the line of sight.}
		\label{fig:nuccioplot}
	\end{figure}

    Let us consider a Cartesian orthogonal reference frame having its origin in the barycentre of the star O and the Z axis along the stellar spin axis. The XY plane is defined in such a way as to contain the line of sight, indicated as $\hat{z}$ (see Figure \ref{fig:nuccioplot}). 
 
    Indicating the latitude and the longitude of a point P on the surface of the star with ($\phi$,$\lambda$), it has the Cartesian coordinate

    \begin{equation}
	P \equiv (X,Y,Z) = R_{\star}(\cos \phi \cos \lambda, \cos \phi \sin \lambda, \sin \phi)
    \end{equation}

    where $R_{\star}$ is the radius of the star assumed to be spherically symmetric. In our fixed reference frame, the longitude $\lambda$ is increasing versus the time $t$ because of the stellar rotation. It changes according to 
    \begin{equation}
	\lambda = \lambda_{\rm 0} + \Omega_{\star}(t-t_{\rm 0})
    \end{equation}
    where $\lambda_{\rm 0}$ is the longitude at the initial time $t_{\rm 0}$ and $\Omega_{\star} \equiv 2\pi/P_{\rm rot}$ is the stellar angular velocity of rotation with $P_{\rm rot}$ the rotation period. 

    To obtain the coordinates in the reference frame adopted in this work, we first make a rotation of the angle $i$ around the $Y$ axis, where $i$ is the inclination of the stellar spin to the line of sight $z$. This brings the $XY$ plane in the plane of the sky which is the same plane of the stellar disc. The equations of such a rotation are
    
    \begin{align}
    \centering
	x &= X \cos i - Z \sin i \\
	y &= Y \\
	z &= X \sin i + Z \cos i
    \end{align}

    This gives
    \begin{align}
	x &= R_{\star}(\cos \phi \cos i \cos \lambda - \sin \phi \sin i) \\
	y &= R_{\star}\cos \phi \sin \lambda \\
	z &= R_{\star}(\cos \phi \sin i \cos \lambda + \sin \phi \cos i)
    \end{align}

    The unit vector along the spin axis of the star in the $OXYZ$ reference frame is $\hat{\Omega_{\star}} \equiv \hat{Z} \equiv (0,0,1)$. Transforming it to the $Oxyz$ reference frame, it becomes $(-\sin{i}, 0, \cos{i})$. Since the $z$ axis is directed along the line of sight, the projection of the spin axis onto the plane of the sky $xy$, that is, the plane of the stellar disc, is $(-\sin{i}, 0)$. The reference frame adopted in the model to describe the disc of the star has the $x_{\rm 0}$ axis orthogonal to the projection of the spin axis on the plane of the sky and the $y_{\rm 0}$ axis along that projection. This implies that our $x_{\rm 0} = y$ and $y_{\rm 0} = -x$ because the projection of the spin axis in the plane of the stellar disc is opposite to the orientation of our x-axis, given that its component along our $x$ axis is $- \sin{i} < 0$ for $0^{\circ} \leq i \leq 180^{\circ}$. 

    In conclusion, we find 
    \begin{align}
	x_{\rm 0} &= R_{\star} \cos \phi \sin \lambda \\
	y_{\rm 0} &= R_{\star}(\sin \phi \sin i - \cos \phi \cos i \cos \lambda)
    \end{align}
    \clearpage
   \onecolumn

    \section{Spot distribution }\label{appendice:2vs3spots}

    \begin{figure*}[ht!]
	\centering
	\includegraphics[width=0.9\hsize]{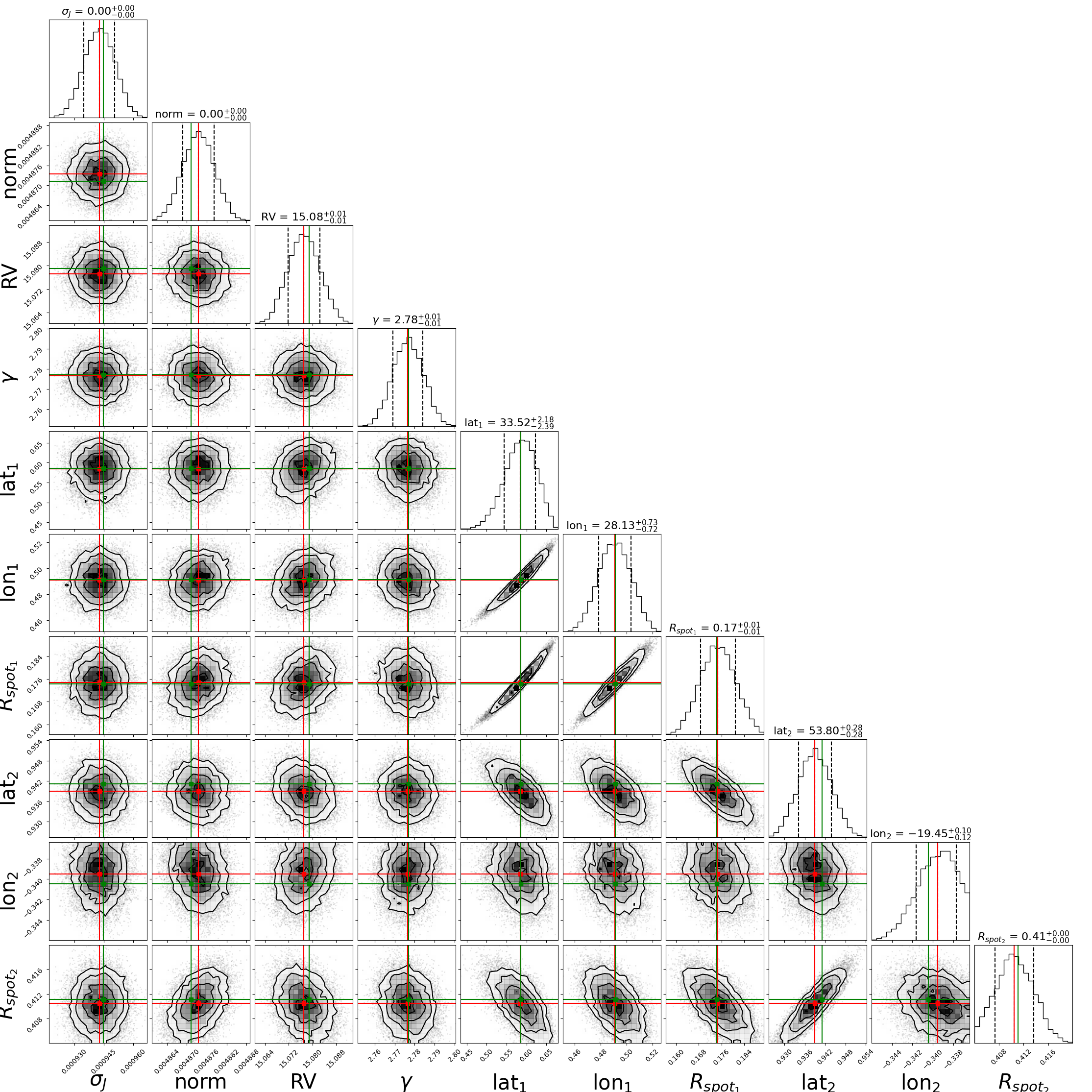}
	\caption{Example of corner plot of the best-fit parameters obtained from the fitting of a CCF profile of V1298 Tau with the "Two-spots model". The red and green lines mark the median and the maximum-a-posteriori values, respectively, while the dashed black lines are the 16$^{\mathrm{th}}$ and 84$^{\mathrm{th}}$ quantiles. The median $\pm$ 1$\sigma$ values are reported in the title of each histogram. The latitude and longitude scales are in radians. log $\mathcal{Z}$ = 31005}
	\label{fig:examplecorner2spot}
\end{figure*}

\begin{figure*}[ht!]
	\centering
	\includegraphics[width=0.9\hsize]{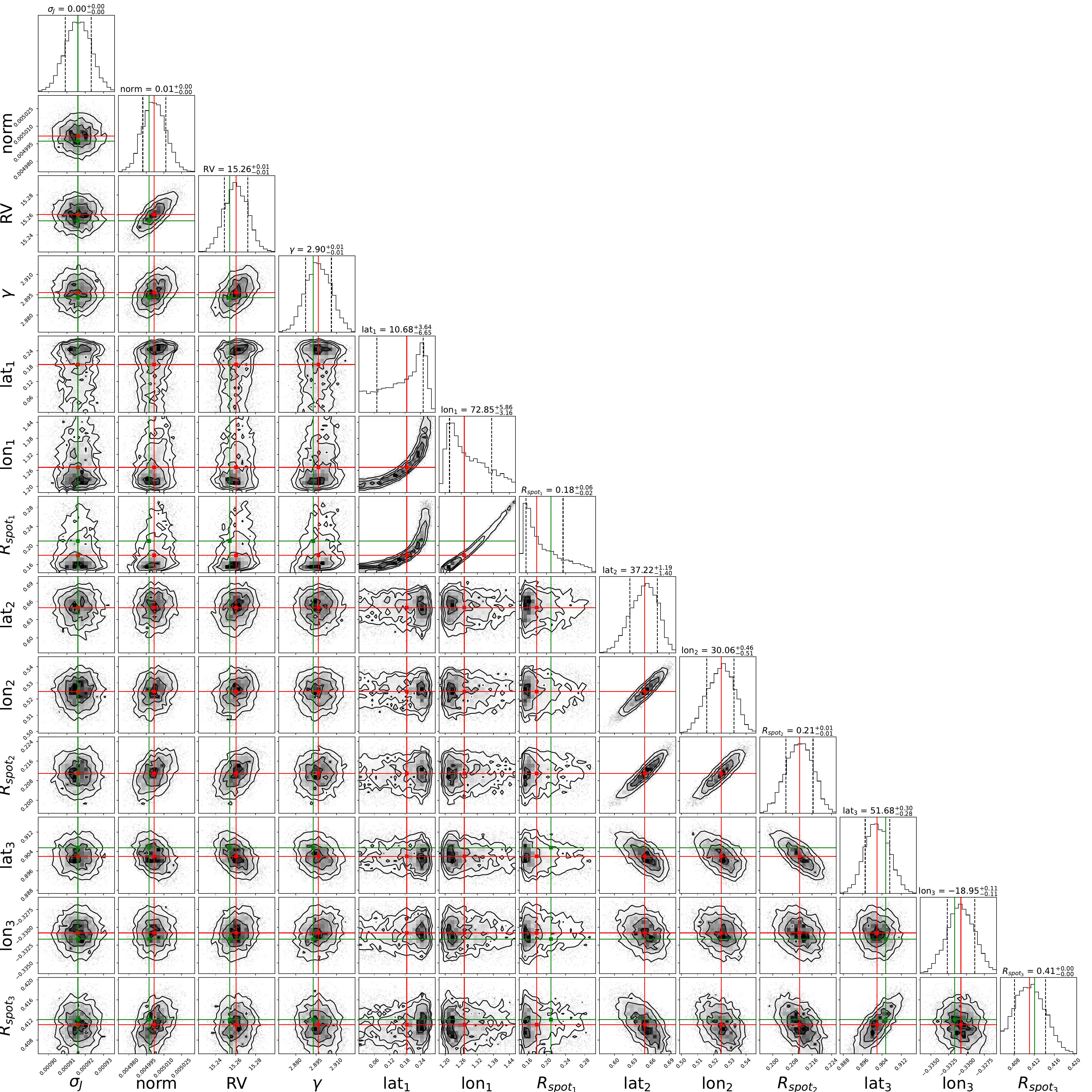}
	\caption{Example of corner plot of the best-fit parameters obtained from the fitting of a CCF profile of V1298 Tau with the "Three-spots model". We can also note that the distributions obtained for the third spot (latitude, longitude and radius) are not well constrained. log $\mathcal{Z}$ = 31156 }
	\label{fig:examplecorner3spot}
\end{figure*}

 \begin{table*}\scriptsize
    \caption{Spot parameters derived by the Two-spots model.}
    \label{tab:params}
    \centering
    \begin{tabular}{cccccccccc} 
    \toprule[0.05cm]
    \toprule
    \multicolumn{10}{c}{\textsc{Spot Parameters}}\\
    \midrule[0.05cm]
    Time (BJD - 2450000) & lat$_1$ (deg)  & lon$_1$ (deg) & R$_{{\rm spot}_{1}}$ (R$_{\star}$) & ff$_{p_{1}}$& lat$_2$ (deg) & lon$_2$ (deg)& R$_{{\rm spot}_{2}}$ (R$_{\star}$) & ff$_{p_{2}}$ & ff$_{p_{\rm tot}}$\\
    \midrule
    8544.3983	&	19	$^{	6	}_{	4	}$	&	23.9	$^{	0.7	}_{	0.7	}$	&	0.149	$^{	0.005	}_{	0.005	}$	&	0.023	&	61.2	$^{	0.2	}_{	0.3	}$	&	-26.6	$^{	0.3	}_{	0.3	}$	&	0.395	$^{	0.004	}_{	0.004	}$	&	0.055	&	0.028	\\
8545.3975	&	23.5	$^{	1	}_{	0.8	}$	&	14.92	$^{	0.09	}_{	0.09	}$	&	0.196	$^{	0.002	}_{	0.001	}$	&	0.041	&	47.6	$^{	0.3	}_{	0.3	}$	&	-13.6	$^{	0.2	}_{	0.2	}$	&	0.246	$^{	0.001	}_{	0.002	}$	&	0.040	&	0.041	\\
8547.3495	&	50.9	$^{	0.8	}_{	0.7	}$	&	52.4	$^{	1.1	}_{	1.2	}$	&	0.271	$^{	0.011	}_{	0.012	}$	&	0.021	&	59.6	$^{	0.2	}_{	0.2	}$	&	-20.54	$^{	0.15	}_{	0.15	}$	&	0.442	$^{	0.003	}_{	0.002	}$	&	0.081	&	0.042	\\
8548.3465	&	56.3	$^{	0.2	}_{	0.2	}$	&	12.01	$^{	0.11	}_{	0.11	}$	&	0.498	$^{	0.003	}_{	0.002	}$	&	0.125	&	52.1	$^{	0.3	}_{	0.4	}$	&	-69.2	$^{	0.4	}_{	0.4	}$	&	0.4990	$^{	0.0020	}_{	0.0010	}$	&	0.046	&	0.126	\\
8559.3780	&	50.3	$^{	0.6	}_{	0.5	}$	&	-43.9	$^{	0.5	}_{	0.5	}$	&	0.271	$^{	0.005	}_{	0.005	}$	&	0.028	&	38.4	$^{	0.9	}_{	0.9	}$	&	5.36	$^{	0.09	}_{	0.09	}$	&	0.234	$^{	0.003	}_{	0.003	}$	&	0.047	&	0.059	\\
8561.3474	&	64	$^{	0.2	}_{	0.2	}$	&	-16.6	$^{	0.3	}_{	0.3	}$	&	0.498	$^{	0.002	}_{	0.001	}$	&	0.090	&	62.9	$^{	0.2	}_{	0.2	}$	&	35.0	$^{	0.4	}_{	0.4	}$	&	0.5000	$^{	0.0020	}_{	0.0010	}$	&	0.079	&	0.090	\\
8562.3467	&	21	$^{	2	}_{	2	}$	&	14.3	$^{	0.2	}_{	0.2	}$	&	0.197	$^{	0.003	}_{	0.003	}$	&	0.043	&	50.0	$^{	0.3	}_{	0.3	}$	&	-31.6	$^{	0.2	}_{	0.2	}$	&	0.362	$^{	0.003	}_{	0.003	}$	&	0.065	&	0.205	\\
8563.3437	&	3	$^{	5	}_{	2	}$	&	27.3	$^{	0.2	}_{	0.3	}$	&	0.09	$^{	0.001	}_{	0.001	}$	&	0.009	&	71.1	$^{	0.2	}_{	0.2	}$	&	-13.6	$^{	0.4	}_{	0.3	}$	&	0.500	$^{	0.001	}_{	0.001	}$	&	0.066	&	0.014	\\
8564.3568	&	57.9	$^{	1.2	}_{	1.2	}$	&	35.1	$^{	1.3	}_{	1.3	}$	&	0.291	$^{	0.013	}_{	0.015	}$	&	0.029	&	25	$^{	3	}_{	3	}$	&	-12.9	$^{	0.2	}_{	0.2	}$	&	0.183	$^{	0.005	}_{	0.004	}$	&	0.036	&	0.040	\\
8565.3427	&	34	$^{	2	}_{	2	}$	&	28.2	$^{	0.7	}_{	0.7	}$	&	0.174	$^{	0.006	}_{	0.006	}$	&	0.024	&	53.9	$^{	0.3	}_{	0.3	}$	&	-19.51	$^{	0.12	}_{	0.10	}$	&	0.411	$^{	0.003	}_{	0.003	}$	&	0.086	&	0.024	\\
8566.3420	&	68.2	$^{	0.2	}_{	0.2	}$	&	-14.1	$^{	0.3	}_{	0.3	}$	&	0.474	$^{	0.004	}_{	0.004	}$	&	0.066	&	5	$^{	6	}_{	5	}$	&	38.1	$^{	0.6	}_{	0.9	}$	&	0.077	$^{	0.002	}_{	0.002	}$	&	0.005	&	0.066	\\
... & ... & ... & ... & ... & ... & ... & ... & ... & ... \\
     \bottomrule[0.05cm]
    \end{tabular}
    \tablefoot{The full table is available at the CDS.}
    \end{table*}

\clearpage
   
   \section{Spot configuration for the same night observations}

    \begin{figure*}[hp!]
	\centering
	{\subfloat[$\Delta$T = 2.53 h\label{fig:spot1A}]{\includegraphics[width=0.45\hsize]{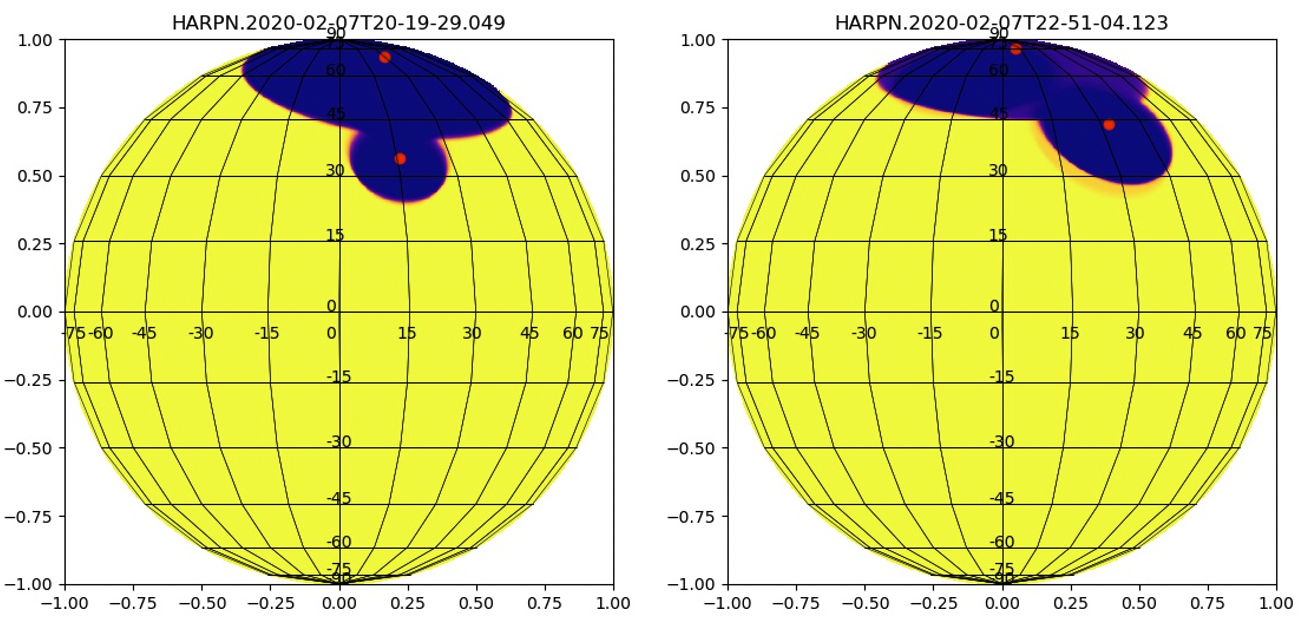}} 
		\subfloat[$\Delta$T = 1.97 h\label{fig:spot2A}]{\includegraphics[width=0.45\hsize]{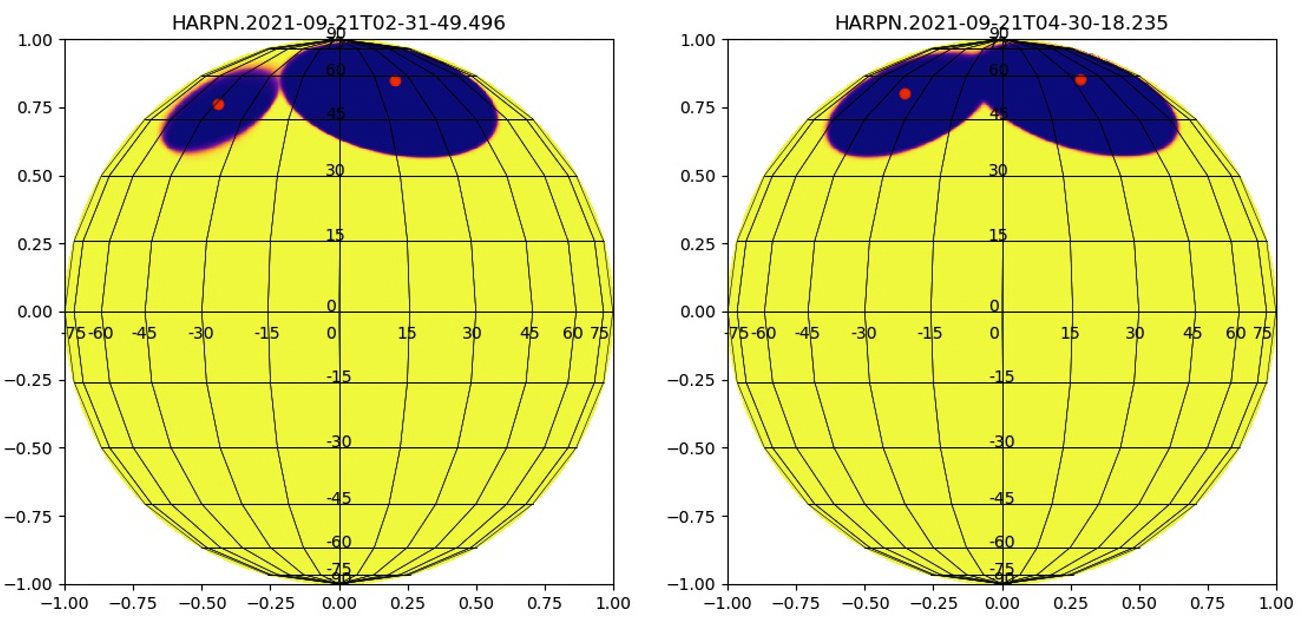}} \\
		\subfloat[$\Delta$T = 1.58 h\label{fig:spot3A}]{\includegraphics[width=0.45\hsize]{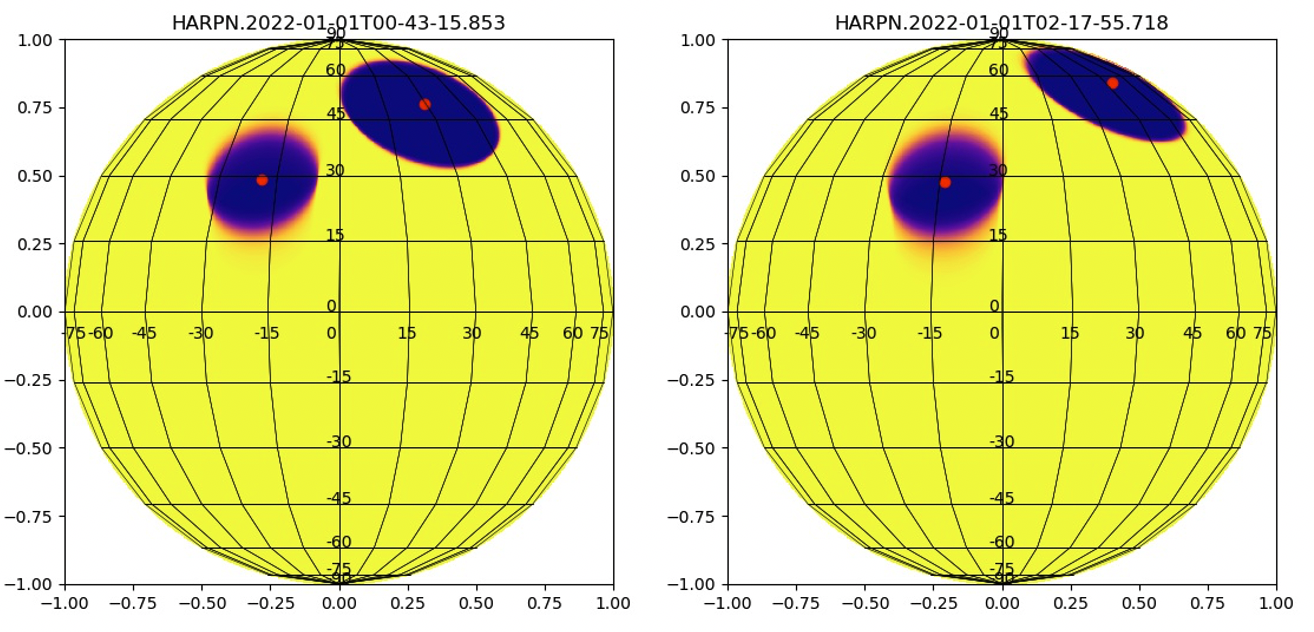}} 
		\subfloat[$\Delta$T = 1.69 h\label{fig:spot4A}]{\includegraphics[width=0.45\hsize]{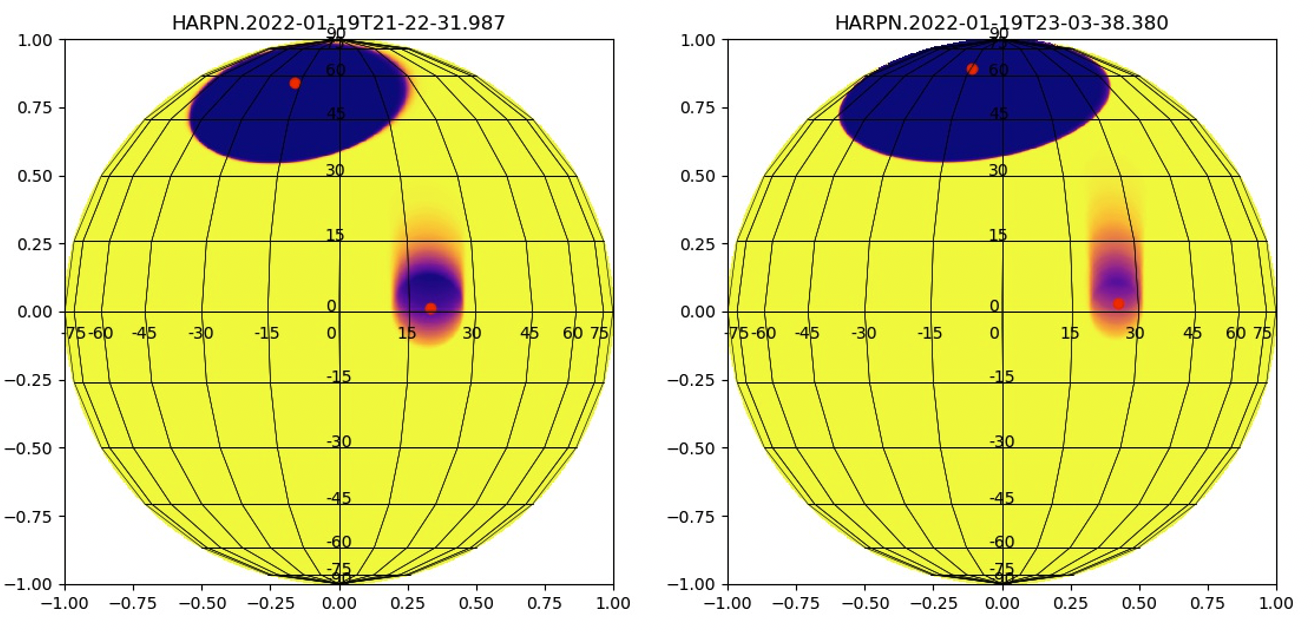}} \\
		\subfloat[$\Delta$T = 3.06 h\label{fig:spot5A}]{\includegraphics[width=0.45\hsize]{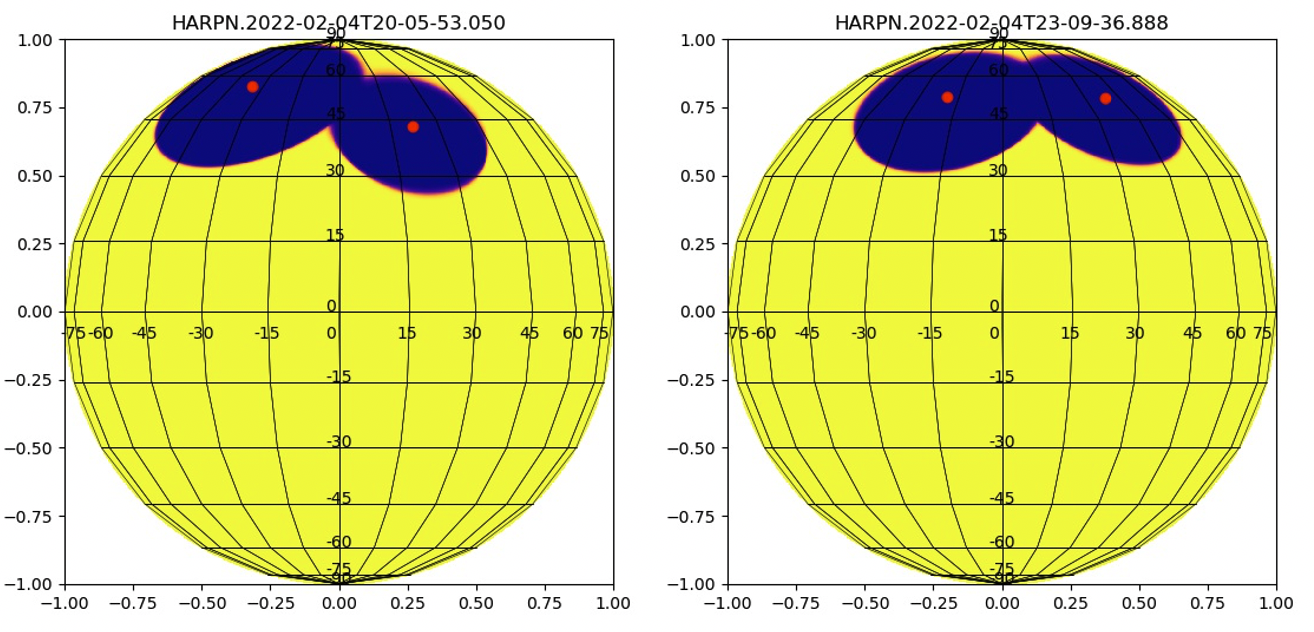}}
		\subfloat[$\Delta$T = 2.20 h\label{fig:spot6A}]{\includegraphics[width=0.45\hsize]{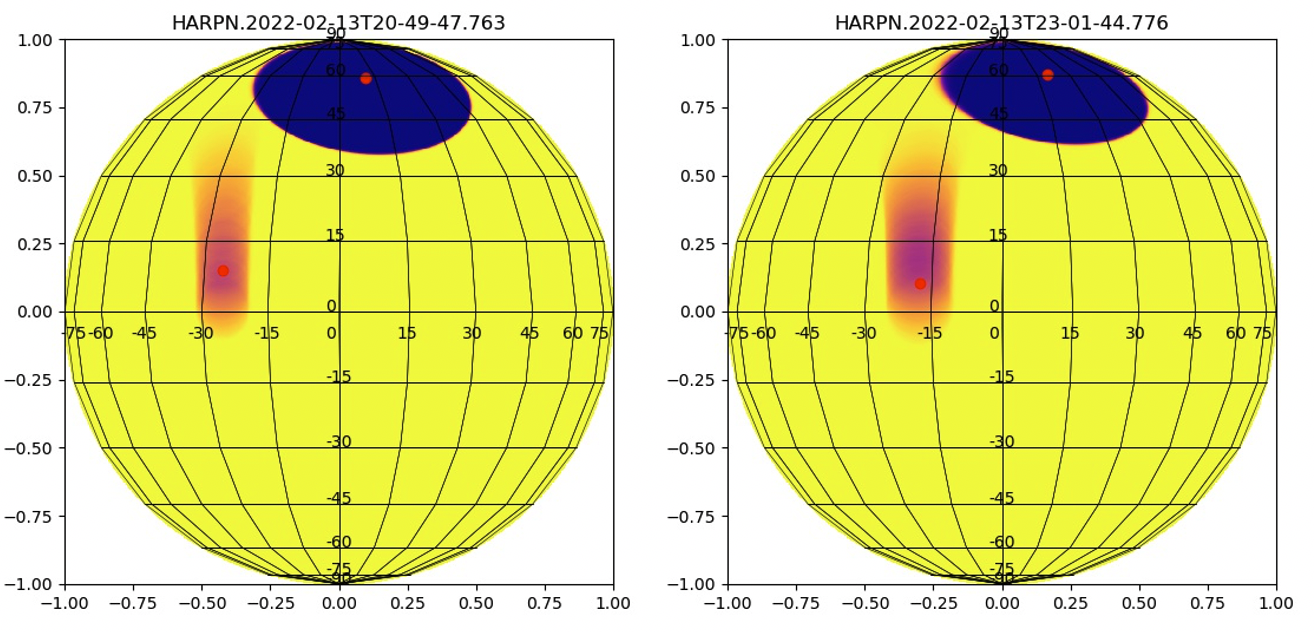}}\\
		\subfloat[$\Delta$T = 2.24 h\label{fig:spot7A}]{\includegraphics[width=0.45\hsize]{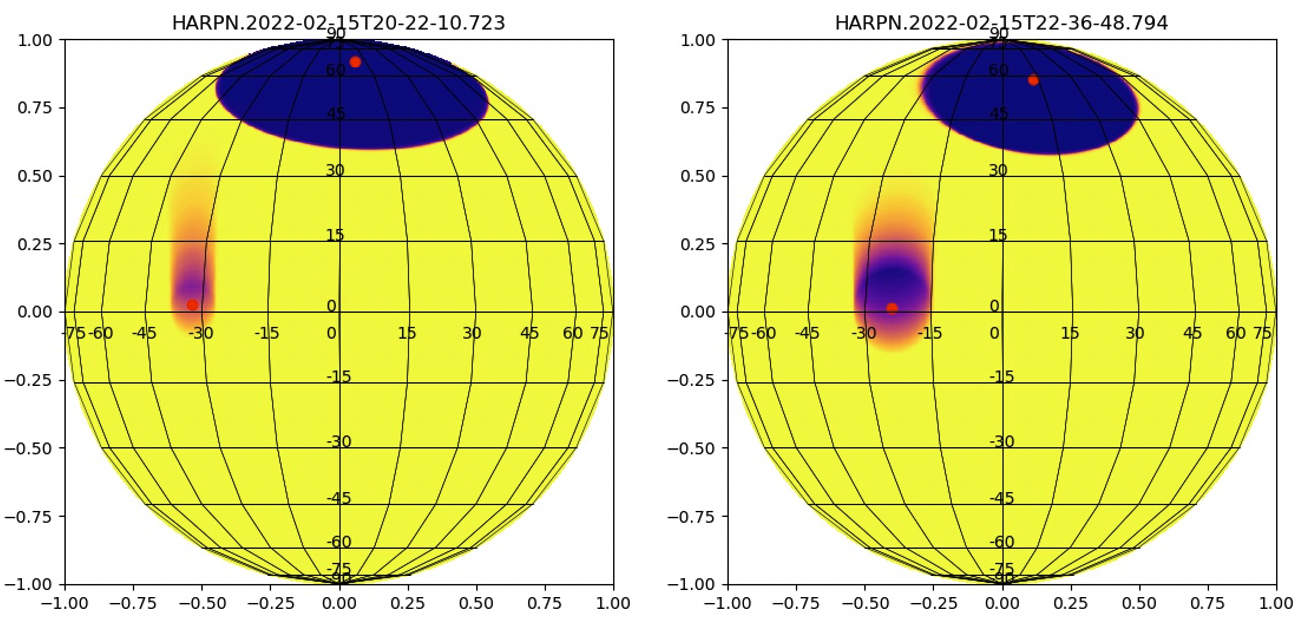}}
		\subfloat[$\Delta$T = 2.45 h\label{fig:spot8A}]{\includegraphics[width=0.45\hsize]{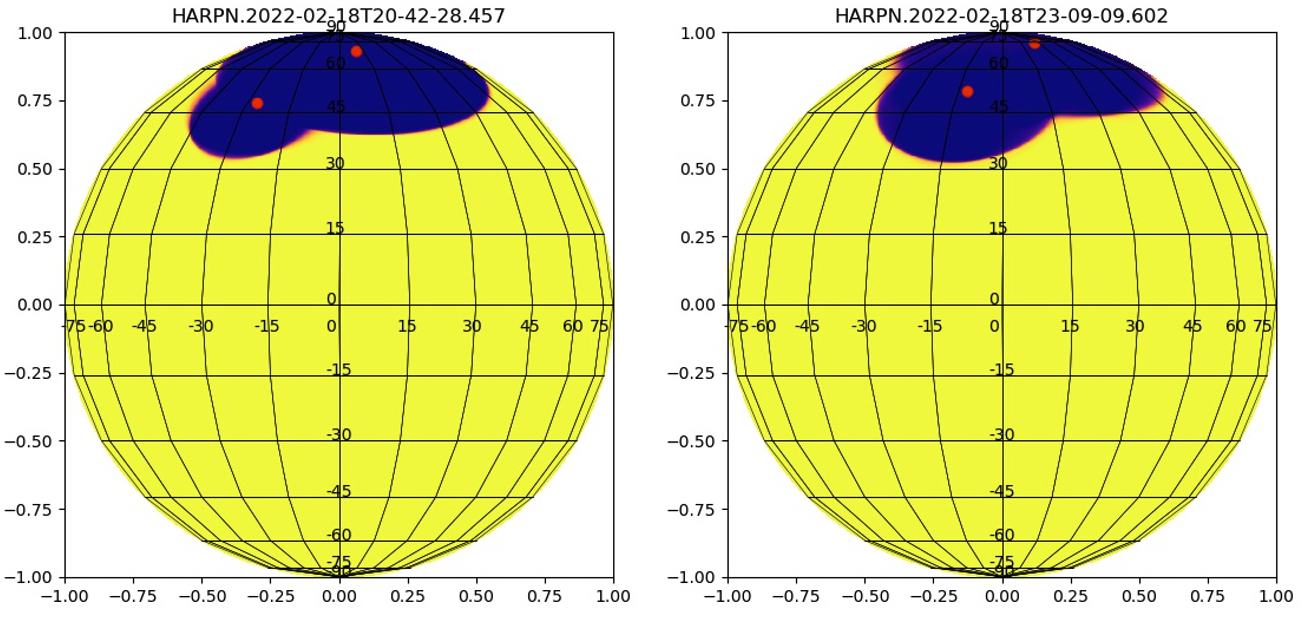}}\\
	}
\caption{Spot configuration of pairs of observations obtained at a few hours of distance. Each map shows the results obtained by the chain. 
At each step, the chain generates a map for each triplet of values (latitude, longitude, and radius), drawing the corresponding spot on the stellar surface. The final map, as well as those shown in this figure, is obtained by summing the individual maps obtained at each step of the chain, using a colour scale ranging from red to blue. The bluest area on the map represents the final configuration of each spot determined by the chain. } 
\label{fig:spots}
\end{figure*}

\begin{figure*}[hp!]
	\ContinuedFloat
	\centering
			{\subfloat[$\Delta$T = 2.16 h\label{fig:spot9A}]{\includegraphics[width=0.45\hsize]{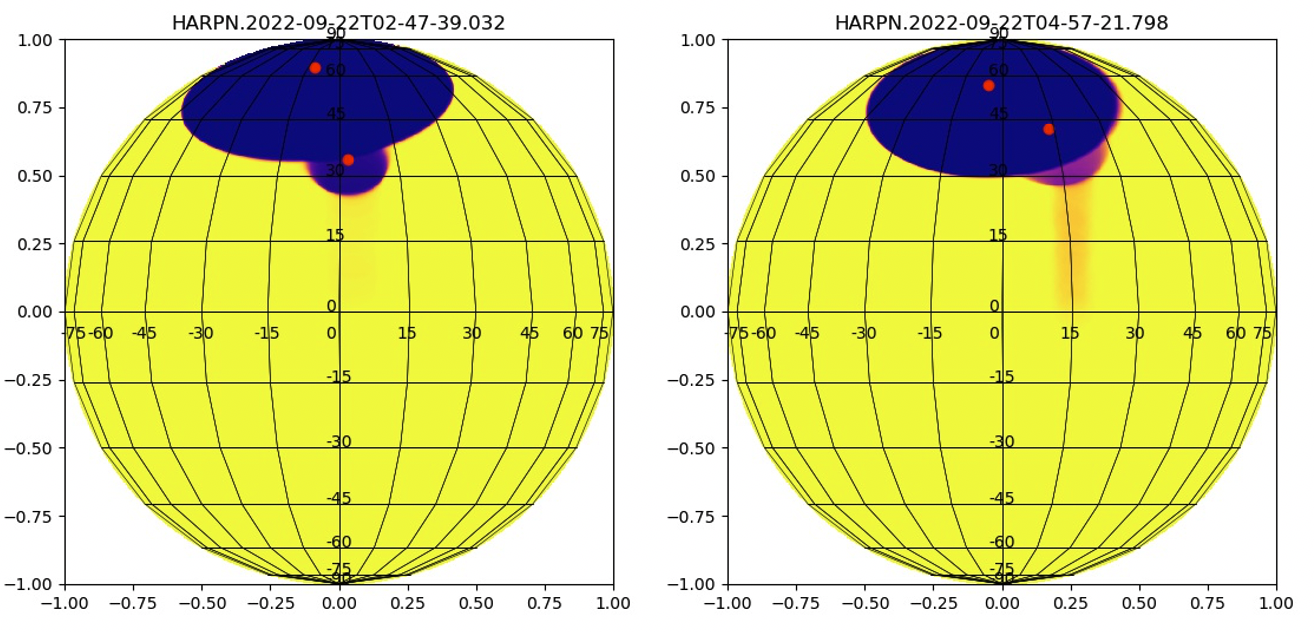}}
		\subfloat[$\Delta$T = 2.79 h\label{fig:spot10A}]{\includegraphics[width=0.45\hsize]{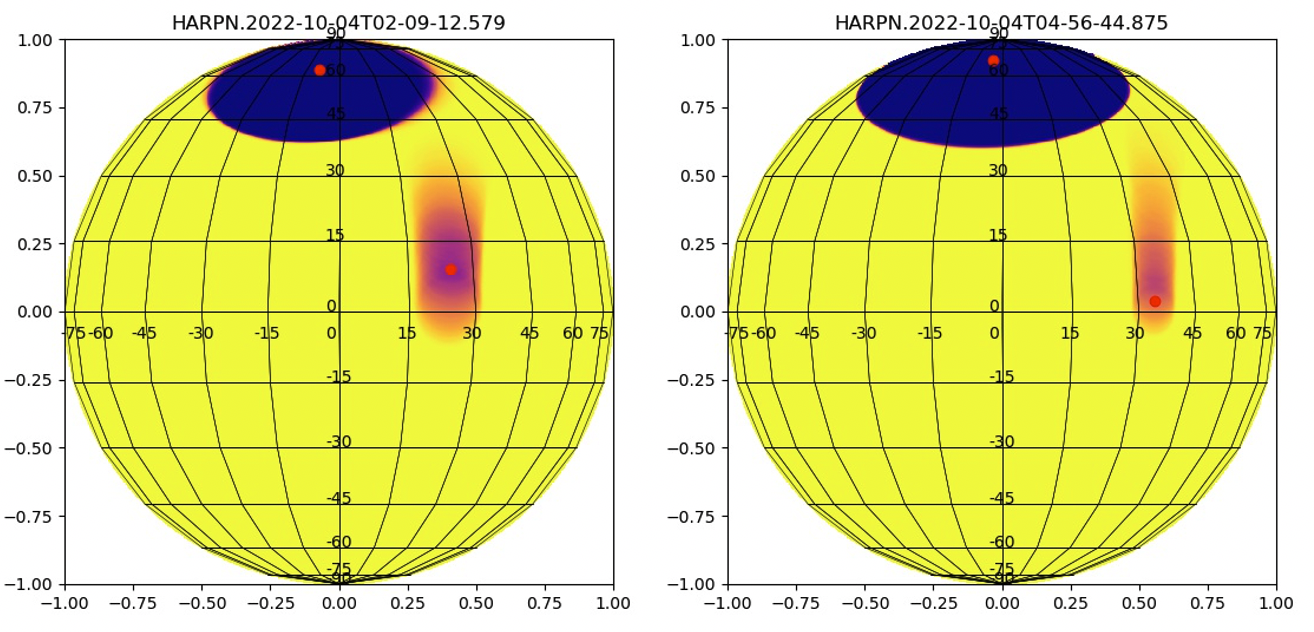}}\\
			\subfloat[$\Delta$T = 1.53 h\label{fig:spot11A}]{\includegraphics[width=0.45\hsize]{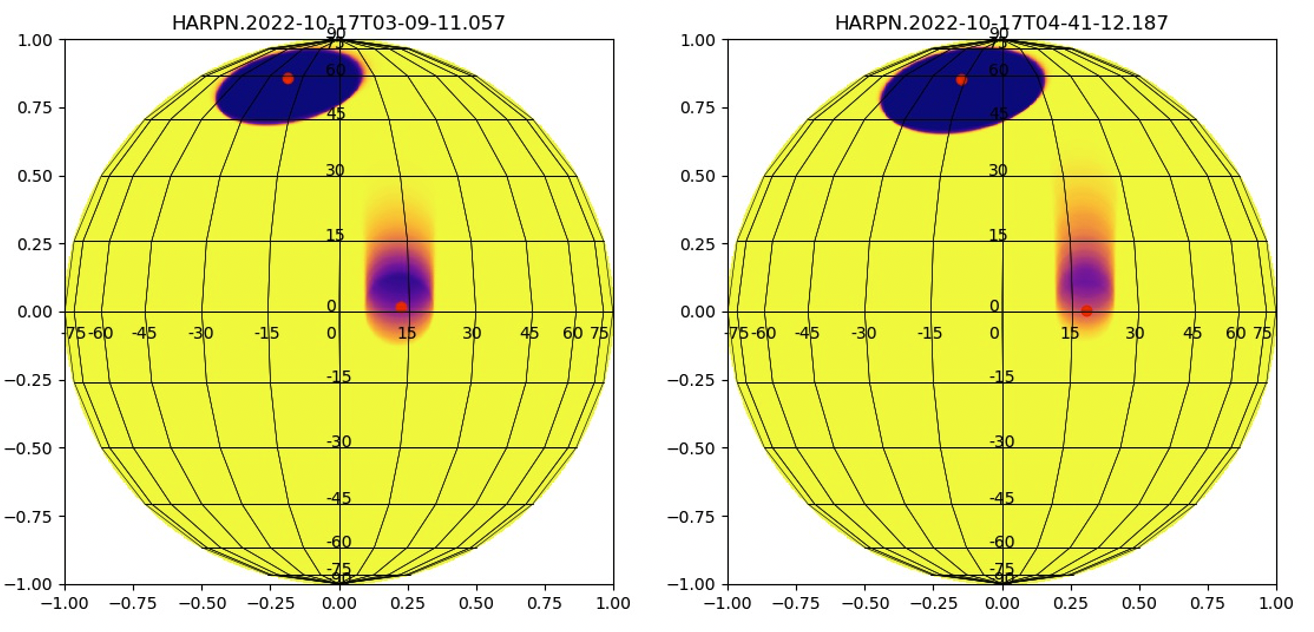}}
		\subfloat[$\Delta$T = 2.74 h\label{fig:spot12A}]{\includegraphics[width=0.45\hsize]{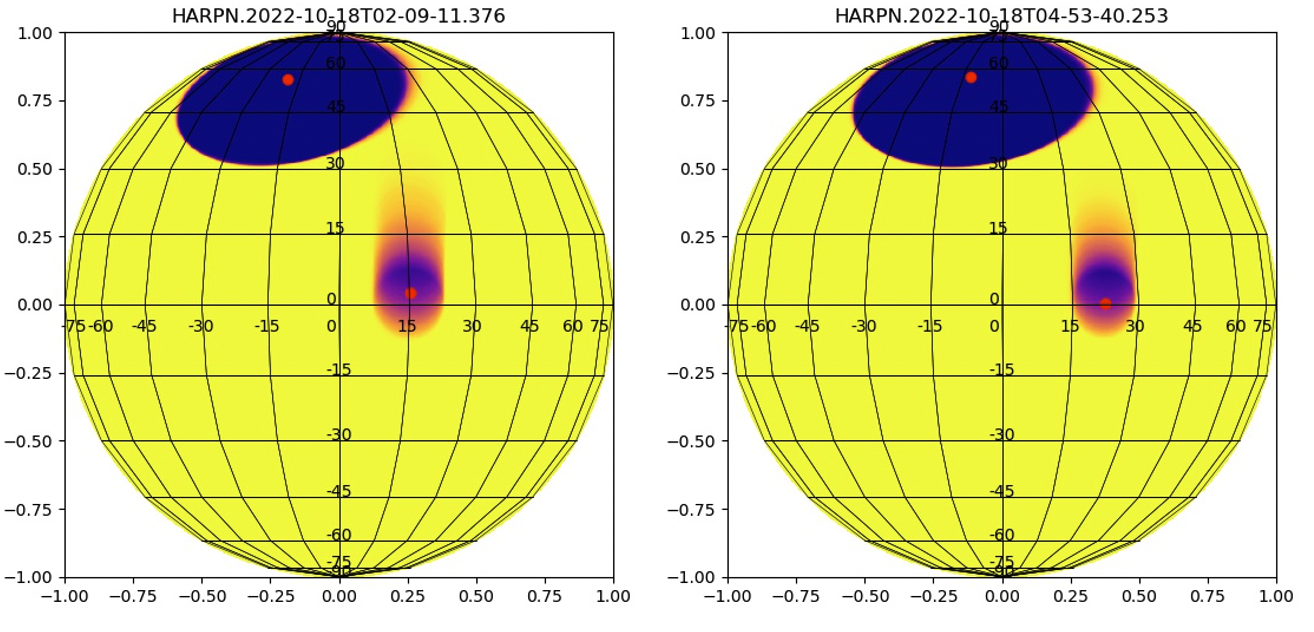}}\\
			\subfloat[$\Delta$T = 2.13 h\label{fig:spot13A}]{\includegraphics[width=0.45\hsize]{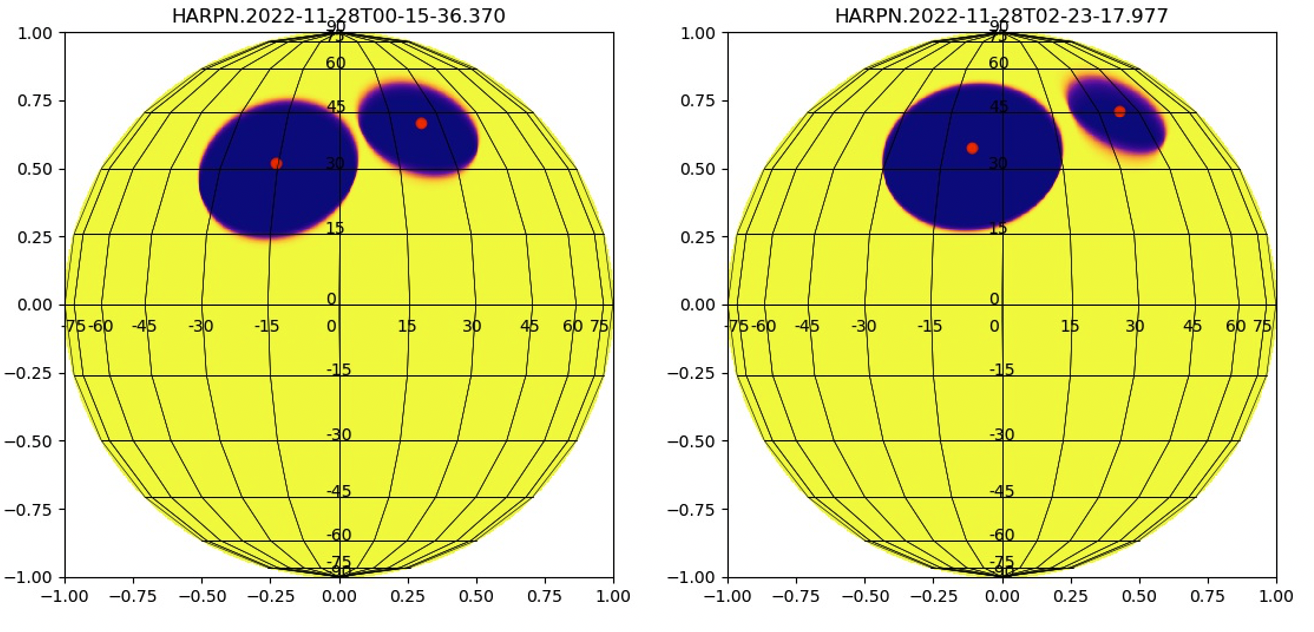}}
		\subfloat[$\Delta$T = 2.70 h\label{fig:spot14A}]{\includegraphics[width=0.45\hsize]{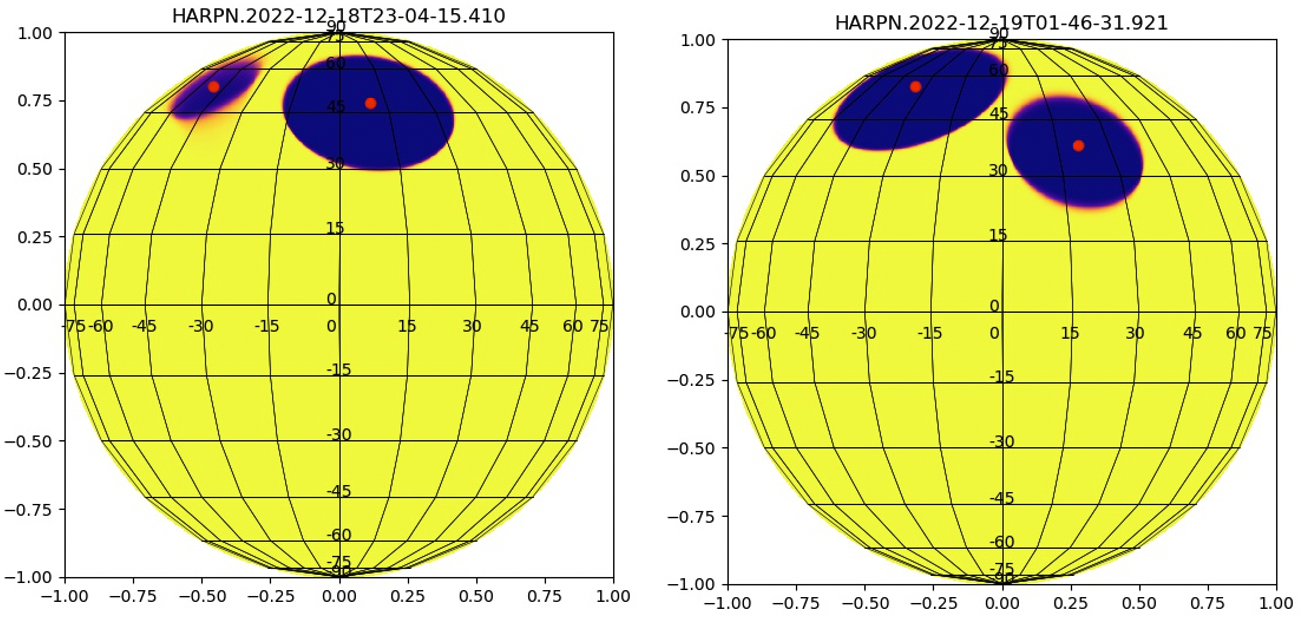}}\\
			\subfloat[$\Delta$T = 1.44 h\label{fig:spot15A}]{\includegraphics[width=0.45\hsize]{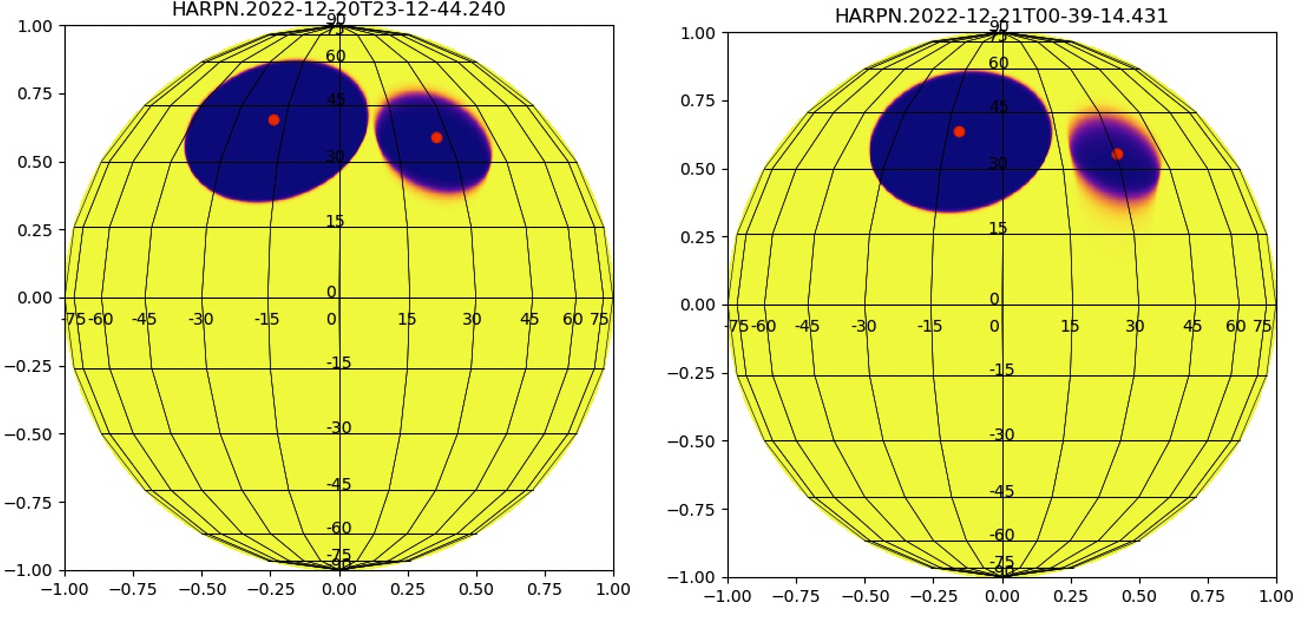}}
		\subfloat[$\Delta$T = 2.00 h\label{fig:spot16A}]{\includegraphics[width=0.45\hsize]{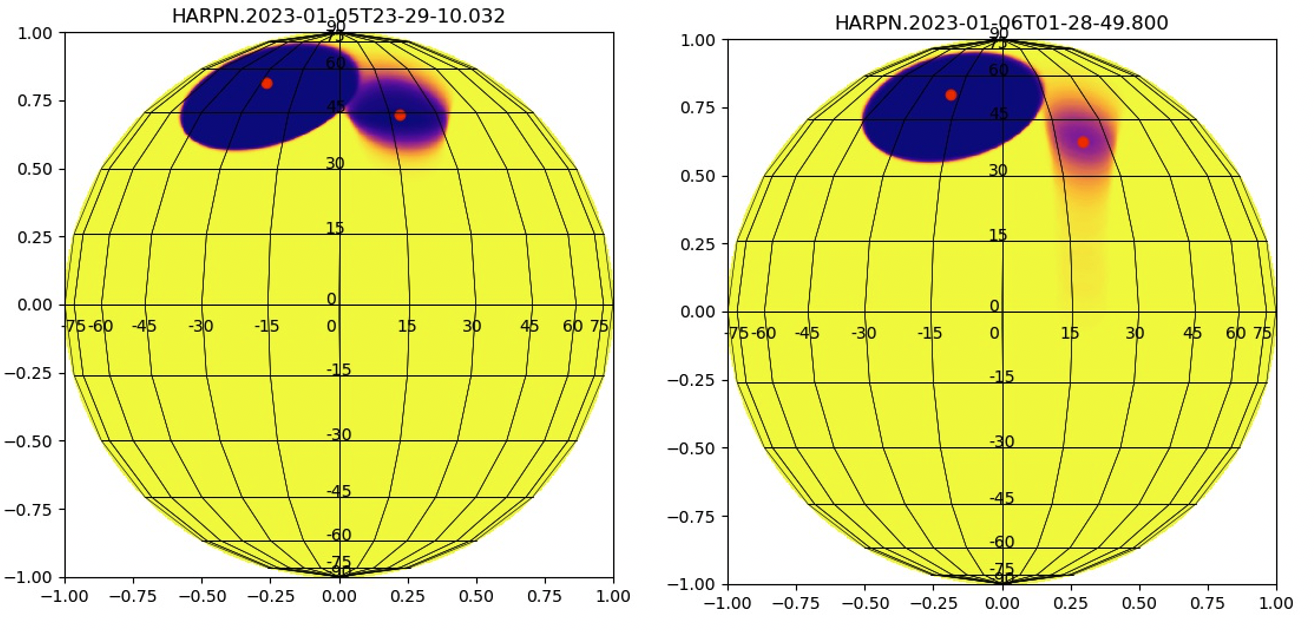}}\\
			\subfloat[$\Delta$T = 2.20 h\label{fig:spot17A}]{\includegraphics[width=0.45\hsize]{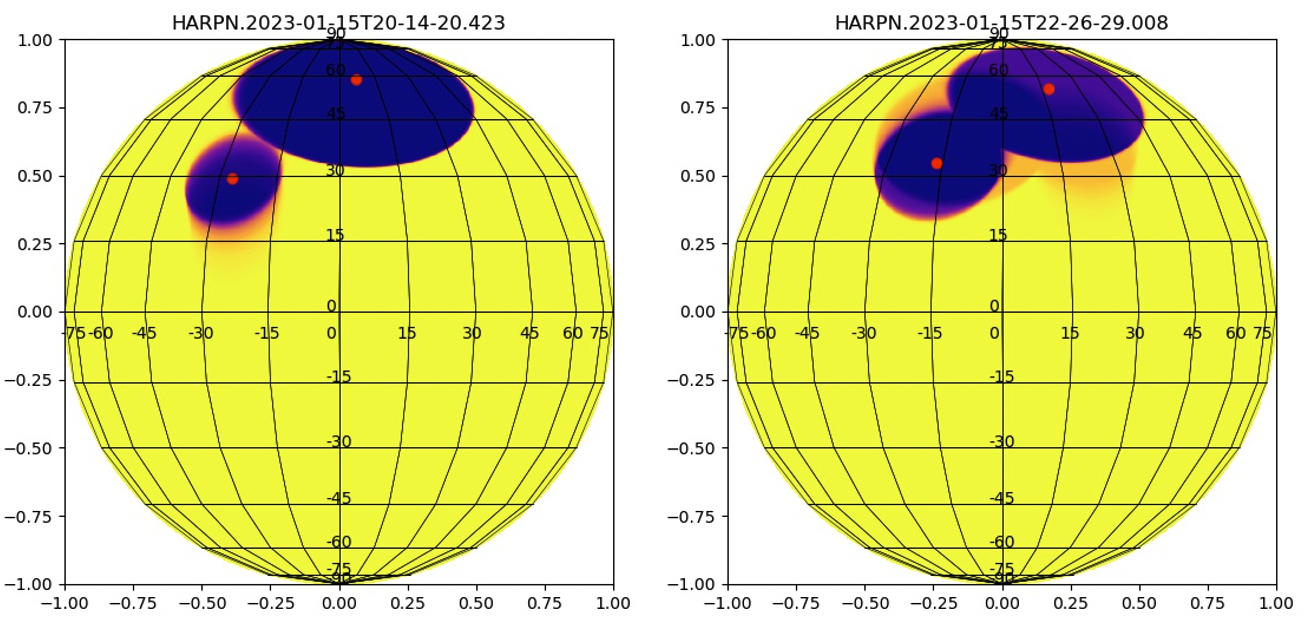}}
		\subfloat[$\Delta$T = 5.09 h\label{fig:spot18A}]{\includegraphics[width=0.45\hsize]{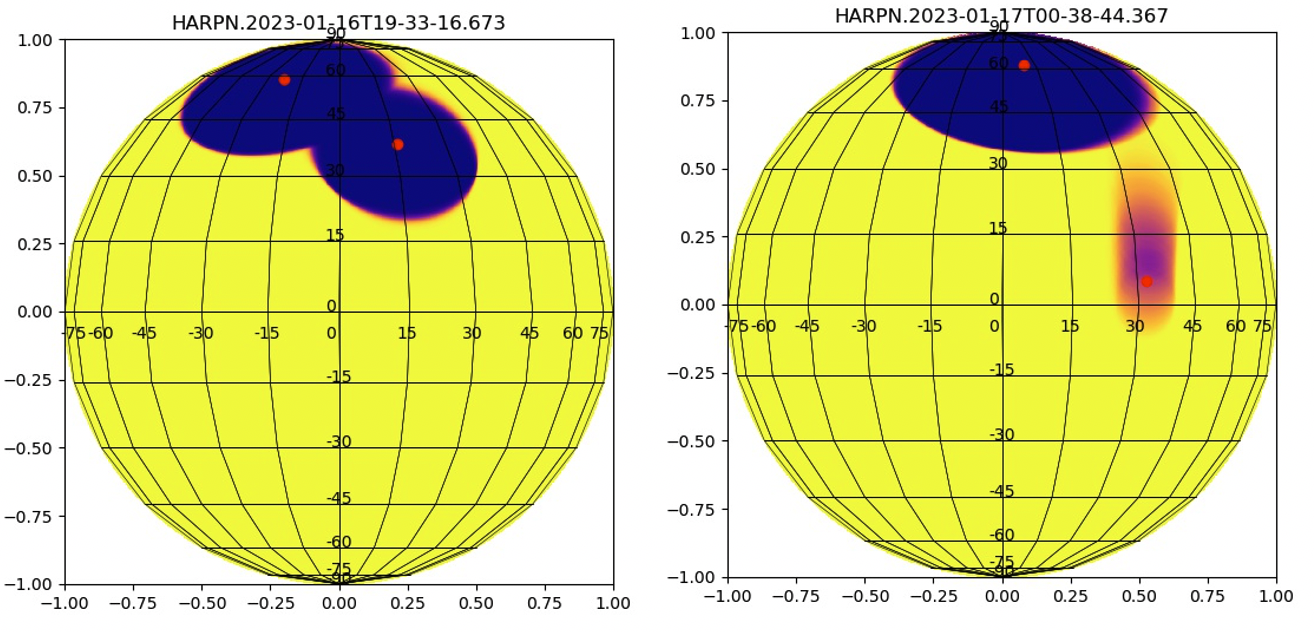}}\\
	}
	\caption{Spot configuration of pairs of observations obtained at a few hours of distance (continued).} 
	\label{fig:spots_continuazione}
\end{figure*}

  \end{appendix}
\end{document}